\documentclass[10pt,twocolumn,letterpaper,journal]{IEEEtran}
\usepackage{graphicx}
\usepackage{amssymb}
\usepackage{amsmath}
\usepackage{multirow}
\usepackage{comment}
\usepackage{algorithm}
\usepackage[]{algpseudocode}
\usepackage{balance}
\usepackage{mathtools}
\usepackage{bm}
\usepackage{url}
\usepackage{cleveref,cite} %
\usepackage{xspace} %

\usepackage{subcaption}
\usepackage[font=footnotesize]{caption}

\usepackage[T1]{fontenc} %

\usepackage{colortbl}

\let\OLDthebibliography\thebibliography
\renewcommand\thebibliography[1]{
 \OLDthebibliography{#1}
 \setlength{\parskip}{0pt}
 \setlength{\itemsep}{5pt plus 0.3ex}
}

\makeatletter
\IEEEtriggercmd{\reset@font\normalfont\fontsize{7.9pt}{8.40pt}\selectfont}
\makeatother
\IEEEtriggeratref{1}

\makeatletter
\def\footnoterule{\kern-3\p@
  \hrule \@width 0.4\textwidth \kern 2.6\p@} %
\makeatother

\IEEEoverridecommandlockouts

\newcommand{\be}{\begin{equation}}
\newcommand{\ee}{\end{equation}}

\begin{document}
\title{Energy-efficient Analog Sensing for Large-scale and High-density Persistent Wireless Monitoring}
\author{Vidyasagar~Sadhu,~\IEEEmembership{Student~Member,~IEEE,} Xueyuan~Zhao,~\IEEEmembership{Member,~IEEE,} and~Dario~Pompili,~\IEEEmembership{Senior~Member,~IEEE}%
\thanks{The authors are with the Department of Electrical and Computer Engineering, Rutgers University--New Brunswick, NJ, USA.}
\thanks{Emails:\{vidyasagar.sadhu, xueyuan.zhao, pompili\}@rutgers.edu}
\thanks{A preliminary version of this work is in the \emph{Proc. of the IEEE Conference on Wireless On-demand Network Systems and Services}, Jackson, WY, 2017~\cite{Sadhu2017wons}.}}%

\maketitle
\thispagestyle{empty}

\begin{abstract}
The research challenge of current Wireless Sensor Networks~(WSNs) is to design energy-efficient, low-cost, high-accuracy, self-healing, and scalable systems for applications such as environmental monitoring. Traditional WSNs consist of low density, power-hungry digital motes that are expensive and cannot remain functional for long periods on a single power charge. In order to address these challenges, a \textit{dumb-sensing and smart-processing} architecture that splits sensing and computation capabilities is proposed. Sensing is exclusively the responsibility of analog substrate---consisting of low-power, low-cost all-analog sensors---that sits beneath the traditional WSN comprising of digital nodes, which does all the processing of the sensor data received from analog sensors.
A low-power and low-cost solution for substrate sensors has been proposed using Analog Joint Source Channel Coding~(AJSCC) realized via the characteristics of Metal Oxide Semiconductor Field Effect Transistor~(MOSFET). Digital nodes (receiver) also estimate the source distribution at the analog sensors (transmitter) using machine learning techniques so as to find the optimal parameters of AJSCC that are communicated back to the analog sensors to adapt their sensing resolution as per the application needs. The proposed techniques have been validated via simulations from MATLAB and LTSpice to show promising performance and indeed prove that our framework can support large scale high density and persistent WSN deployment.
\end{abstract}


\section{Introduction}\label{sec:introduction}%

\textbf{Overview:}
Wireless Sensor Networks~(WSNs) are currently used for several purposes
including environmental monitoring~\cite{Sadhu2018ucomms, Rahmati2019Secon},
intelligent transportation systems~\cite{Lu2014},
infrastructure surveillance~\cite{Sadhu2016icac, Chen2014, Sadhu2017percom} 
and other Internet of Things~(IoT) applications~\cite{Kato2017, Mahmud2017, Sadhu2019tmc}. 
Sensors in these networks should be able to capture high spatial and temporal resolution exhibited by the corresponding phenomenon.
Motes in existing/traditional WSNs are composed of digital processors, multiple Analog-to-Digital Converters~(ADCs) and wireless transceivers. These digital motes sense the environment but also carry out digital communications and computations, both of which also require high bit resolution for high precision and dynamic range. Digital motes as a result tend to be power hungry. On the other hand, sensing and basic communications can be carried out by power-efficient analog sensors.

\textbf{Proposed Approach:}
The proposed paradigm breaks away from the past design goal of WSNs comprising high-power, resource-rich digital motes with integrated sensing, communication, and computation capabilities. Instead, a ``dumb-sensing, smart-processing'' architecture (Fig.~\ref{fig:big_picture}) is advocated---comprising a high density of extremely low-power/low-cost \emph{``dumb'' all-analog sensors} with limited communication capabilities as the \emph{sensing substrate} over which exists the ``traditional WSN'' consisting of a low density of \emph{``smart'' digital nodes}---Cluster Heads~(CHs), drones, smart cameras, etc.---with decent communication and computation capabilities, which outsource computation-intensive tasks to the \emph{cloud of cyber-infrastructure} (or simply `cloud') consisting of high-performance computing servers on a need basis (Fig.~\ref{fig:big_picture}).
The \textit{low-cost} factor enables deploying these sensors in \textit{large scale} and \textit{high-density} thereby providing high \textit{spatial accuracy}. The \textit{low-power} on the other hand means the sensors need not be put to sleep thereby providing high \textit{temporal accuracy} unlike the current digital nodes which go to sleep occasionally to conserve power. 
For this, sensor nodes with Shannon-mapping capabilities are envisioned, to enable low power consumption. The Shannon mapping~\cite{Shannon49} is a low-complexity technique for Analog Joint Source-Channel Coding (AJSCC)~\cite{Hekland05}; it can compress two or more signals into one (introducing controlled distortion) while also staying resilient to wireless channel impairments. AJSCC is realized via the current-voltage~(IV) characteristics of a Metal Oxide Semiconductor Field Effect Transistor~(MOSFET). Through this novel approach, power of the order of few tens of $\mu \rm{W}$ (possibility of few $\mu \rm{W}$, as explained in Sect.~\ref{sec:perf_eval}) is achieved for the encoding circuit compared to several tens of $\mu \rm{W}$ in previous works~\cite{Zhao2018a}. This means that a compact energy-harvesting unit~\cite{Chao2011} can power the sensor and enable persistent monitoring; e.g., a tiny solar cell (given the sensor scale of a few $\mathrm{cm^2}$)
or energy harvested from vibrations~\cite{piezodatasheet} (e.g., in case of a bridge) can provide several tens of ${\mu \mathrm{W}}$-level power supply, thus extending the substrate sensors' lifetime to years. This is achieved by working fully in the analog domain and by avoiding power-hungry ADCs and microprocessors, which are used in digital sensing motes. The digital nodes, apart from processing the big data received from substrate sensors, also estimate the source distribution using this data, so as to optimize the encoding parameters of the AJSCC of the substrate sensors as per application needs. As explained later, this is done via minimizing the Kullback-Leibler Divergence~(KLD) score of the estimated distribution with the reference distributions. In order to enable the sensors to receive different configuration information, a circuit design that can accept different configurations based on the information received from digital nodes is proposed.

\textbf{Related Work:}
While the general idea of tiered architectures for WSNs is not new to the research community; see, e.g.,~\cite{Shah.etal.AHN2003,Wang.etal.EJAiSP2003},
the main contribution of this article lies in the introduction of a \textit{sensing substrate} of all-analog sensors with low-power sensing and communication capabilities to enable large-scale wireless monitoring with high spatial and temporal resolutions.
Most of the existing JSCC-hardware solutions are all digital and power hungry. For example, a Software-Defined Radio~(SDR) system to realize AJSCC mapping has been reported in~\cite{Garcia11}. The mapping was also recently implemented in an optical digital communication system in~\cite{Romero14} and has been combined with Compressive Sensing~(CS) in~\cite{Saleh12} to improve robustness against channel noise. For more related work, please refer Table~I in our preliminary work~\cite{Sadhu2017wons}.
There are a couple of works that try to realize JSCC in analog domain like we do. However, these realize the rectangular JSCC unlike ours where we use a novel space-filling curve. To the best of the authors' knowledge, ours is the first work to realize a different space-filling curve in hardware analog domain than the rectangular JSCC. Among those that realize rectangular AJSCC, Zhao et al.~\cite{Zhao2016} proposed all-analog sensor design that realizes AJSCC using Voltage Controlled Voltage Sources~(VCVSs). This design, which they call, ``Design~1'' is an inefficient design as it adopts a fixed number of JSCC levels, hardware in each stage is duplicated, and does not scale with the number of levels. 
Even though the authors proposed ``Design~2''~\cite{Zhao2018a} to address the above limitations, it still has higher power consumption. Design~1 with 11 JSCC levels (quantization levels on y-axis) consumes $130~\rm{\mu W}$, whereas Design~2 with 16 levels consumes $72~\rm{\mu W}$ ($64~\rm{\mu W}$ for 8 levels). These numbers, which do not include the transmission power, are large for sensors powered using energy-harvesting techniques that produce only tens of $\mu \rm{W}$~\cite{Khan2018} to power the entire sensing/transmitting device. 
Differently from above, we adopt the MOSFET's~IV characteristics as the space-filling curve and are able to achieve encoding power consumption of $\approx 24~\mu\rm{W}$, with possibility of $8~\mu\rm{W}$.

\textbf{Applications and Broader Impacts:}
The broader impacts of our solution include low-cost, high-resolution, high-confidence, persistent, end-to-end wireless monitoring for urban infrastructure, precision agriculture, Intelligent Transportation Systems~(ITS), military surveillance, and smart cities, to name a few, in addition to IoT-based solutions such as Body Area Networks~(BANs). Furthermore, our solution can enable various practical applications in aqueous environments such as Underwater IoT~(UW~IoT) to perform oceanographic data collection, pollution and environmental monitoring, tsunami detection/disaster prevention, assisted navigation, and tactical surveillance among others. Our solution can also enable smart home and smart health applications where the environmental and health monitoring are performed concurrently in the wireless sensors with signal compression capabilities.

\textbf{Our Contributions}
can be summarized as follows:
\begin{itemize}
  \item We introduce a \textit{dumb-sensing} and \textit{smart-processing} framework for WSNs that splits sensing and computational tasks between energy-efficient low cost substrate sensors and the traditional digital nodes respectively. 

  \item We design a Metal Oxide Semiconductor Field Effect Transistor~(MOSFET) based realization of the AJSCC-based sensor encoding technique (with corresponding decoding at the receiver) to realize low-power and low-complexity substrate sensors.

  \item We develop a Kernel Density Estimation~(KDE) and Kullback-Leibler Divergence~(KLD)-based technique to estimate the source distribution at the digital CHs, and thereby optimize the AJSCC parameters ($\phi$) considering the wireless channel and the mapping circuit characteristics. This KDE optimization is verified via simulations. 
  
  \item Other novelty of this work consists in its multidisciplinary nature, i.e., this research cuts across fields of semiconductor components~(MOSFET) and machine learning~(KLD-based KDE) with the background of signal compression by the AJSCC.

\end{itemize}

\textbf{Article Outline:}
In Sect.~\ref{sec:prop_soln}, we present our energy-efficient analog sensing solution for large-scale/high-density persistent wireless monitoring;
in Sect.~\ref{sec:perf_eval}, 
we evaluate our techniques using both MATLAB and LTSpice simulations; finally, in Sect.~\ref{sec:conc}, we draw conclusions and discuss future work.

\section{Proposed Solution}\label{sec:prop_soln}
We
describe our modular sensing architecture in Sect.~\ref{sec:prop_soln:arch}. In Sect.~\ref{sec:prop_soln:ajscc}, we focus on the solutions for analog substrate (transmitter) viz., our MOSFET-based AJSCC encoding. In Sect.~\ref{sec:prop_soln:kldiv}, we focus on solutions for digital nodes/CHs (receiver) where we explain our AJSCC-decoding process, propose KDE and KLD-based algorithms to estimate source distribution, and then find the optimal AJSCC parameters.

\subsection{Modular Sensing Architecture}\label{sec:prop_soln:arch}
\begin{figure}
\begin{center}
\includegraphics[width=3.4in]{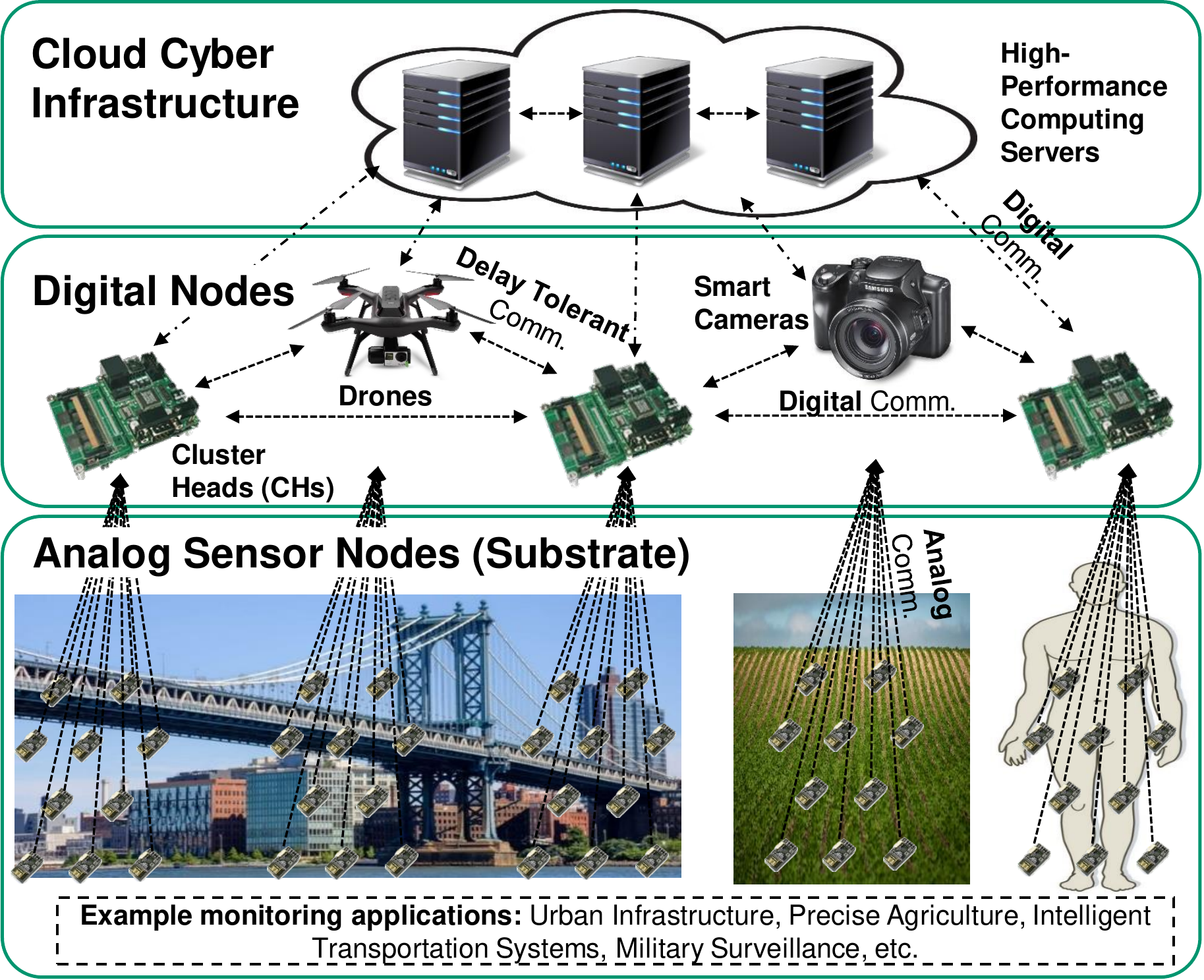}
\end{center}
\vspace{-0.15in}
\caption{
An overview of the proposed architecture.
}\label{fig:big_picture}
\vspace{-0.2in}
\end{figure}

We propose a modular sensing architecture that splits the sensing and computing functionalities between extremely low-cost, low-power analog sensors \emph{and} resource-rich digital nodes and cyber infrastructure, respectively, will allow to \emph{improve} energy efficiency and sensing resolution, \emph{simplify} network management, and \emph{reduce} the cost of network operation, \textit{as the analog sensing substrate can be incrementally modified/added to any existing WSN}. 
An illustration of this vision for high-density sensing is provided in Fig.~\ref{fig:big_picture}, in which sensing is exclusively the responsibility of high-density, miniaturized all-analog sensor nodes (sensing substrate), while processing and computing are exclusively the responsibility of low-density digital nodes/CHs and cloud cyber-infrastructure, respectively.

One of the major hurdles in realizing our vision of an all-analog sensing platform is to develop low-power, low-cost analog sensors.
We will address this challenge by designing and developing wireless transmitting sensors with Shannon-mapping~\cite{Shannon49} capabilities, a low-complexity technique for AJSCC~\cite{Hekland05}.
In agreement with the Latin phrase, \emph{Natura non facit saltus} (``nature does not make jumps"), our sensors are analog, as measurements are taken from the real world, which is inherently analog. An all-analog node should consume, in theory, on the order of tens of $\mu \rm{W}$---which is comparable to the power that can be harvested using, for example, a compact solar panel or a piezoelectric Micro Electro Mechanical Systems~(piezo-MEMS) vibrational energy harvester
---while a Commercial Off-The-Shelf~(COTS) node's power consumption is typically on the order of $\rm{mW}$. 

The next major hurdle in realizing our vision with end-to-end all-analog sensing platforms comes in relation to processing and computing at the digital nodes and cyber infrastructure. For this purpose, the analog signals from multiple sensors will need to be detected and demultiplexed at the digital CHs, which will then process these signals and forward their summary statistics/important features to cyber infrastructure for further processing and possible generation of control commands in ``closed-loop scenarios.'' In a realistic setting, however, digital CHs will have to be programmed to deal with calibration errors, sensor failures, adversarial attacks, and intermittent connectivity to the cloud. while developing techniques to make the CHs resilient to such adversities is outside the scope of this work, we however consider one particular closed-loop scenario in this article---CHs estimate the source distribution so as to optimize the AJSCC encoding parameters at the transmitter and convey that information back to the sensors~(Sect.~\ref{sec:prop_soln:kldiv}).

\subsection{Analog Sensing Substrate (Transmitter)}\label{sec:prop_soln:ajscc}
In this subsection, which deals with the analog sensing substrate (transmitter), we first introduce AJSCC; then, present our novel idea of using MOSFET's characteristics as the space filling curve to realize AJSCC.

\begin{figure}
\begin{center}
\includegraphics[width=3.6in]{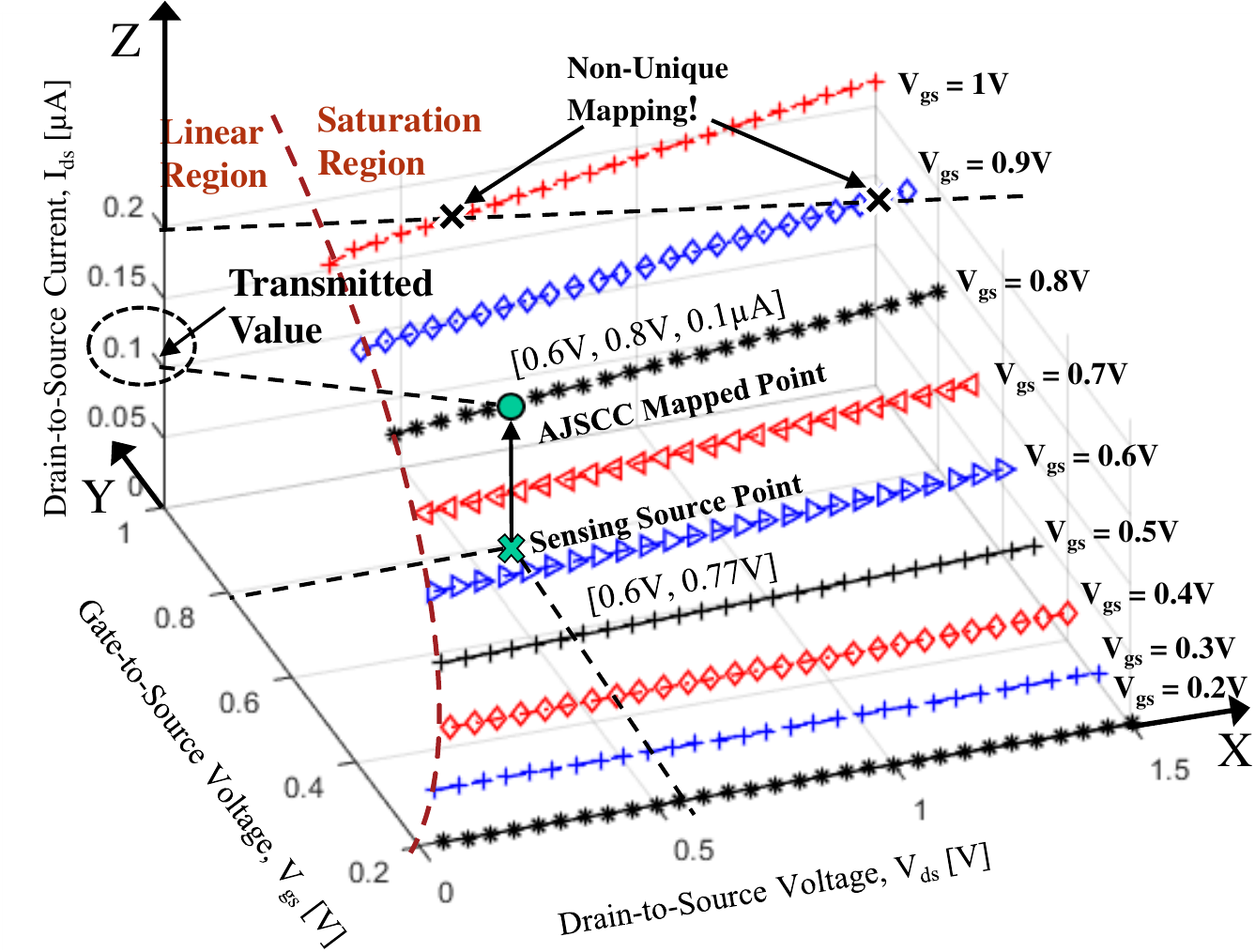}
\end{center}
\vspace{-0.15in}
\caption{Shannon mapping realized via output characteristics ($I_{ds}$ vs. $V_{ds}$ for different $V_{gs}$) of a MOSFET in saturation region (right of dashed line).}\label{fig:ajscc_mos_3d}
\vspace{-0.2in}
\end{figure}

\textbf{(1)~Analog Joint Source Channel Coding~(AJSCC):}
AJSCC is a low-complexity encoding technique, also known as Shannon mapping~\cite{Shannon49}, which compresses two (or more) signals into one. JSCC achieves this using a space-filling curve where the x-axis signal, $x_1$, is continuously captured while the y-axis one, $x_2$, is quantized.
The sensed $(x_1,x_2)$ point is mapped to the closest point on the curve and the encoded (compressed) value is a property of the curve, e.g., length of the curve from origin. To achieve the low-power/low-complexity advantages of JSCC, this technique needs to be realized in the analog domain---hence the name Analog JSCC or AJSCC. However, AJSCC is hard to realize on hardware in a power-efficient manner. This is especially important for our proposed architecture (Fig.~\ref{fig:big_picture}), where the substrate sensors need to be extremely power-efficient so they can be powered using energy-harvesting techniques.

\textbf{(2)~FET-based Encoding at Transmitter:}
To realize JSCC in an energy-efficient manner, we take a completely different path compared to previous approaches that implement rectangular Shannon Mapping~\cite{Zhao2016, Zhao2017}. We realize JSCC using the input-output (also called IV, which stands for current-voltage) characteristics of a single MOSFET device as the space-filling curve for JSCC. Ideally, any new space-filling curves for AJSCC should preserve these properties: ($i$)~they should achieve better trade-off between channel noise/compression and approximation noise; ($ii$)~they should be realizable using \emph{all-analog} components; and ($iii$)~they should result in a \emph{unique} mapping (i.e., two or more sensor values should map to \textit{only one} AJSCC encoded value). Given these desirable properties of a space filling curve, we propose the idea of using the IV characteristics of a MOSFET in saturation region as the space-filling curve (instead of using rectangular parallel lines as used in~\cite{Zhao2016, Zhao2018a}). A MOSFET has three terminals: Gate~(G), Drain~(D), and Source~(S). When a suitable voltage is applied across G and S terminals, $V_{gs}$, and D and S terminals, $V_{ds}$, a current is generated across D and S terminals, $I_{ds}$. The relationship among $V_{gs}$, $V_{ds}$, and $I_{ds}$ for a \emph{real} MOSFET in the saturation region (Fig.~\ref{fig:ajscc_mos_3d}) is, 
\begin{equation} \label{eq:ids_clm}
I_{ds}=\frac{1}{2} \cdot \frac{W}{L} \cdot {\mu}C_{ox} \cdot (V_{gs}-V_{th})^2 \cdot (1+{\lambda} V_{ds}),
\end{equation}
where $W,L~[\rm{m}]$ are width and length of the MOSFET channel, respectively, $\mu~[\rm{m^2/Vs}]$ is the electron mobility in the channel, $C_{ox}~[\rm{F/m^2}]$ is the oxide capacitance per unit area, and $\lambda~[\rm{V^{-1}}]$ is the Channel Length Modulation~(CLM) parameter. Because of CLM, $I_{ds}$ keeps increasing at a very slow rate (governed by $V_{gs}$ and other parameters) in the saturation region.

Fig.~\ref{fig:ajscc_mos_3d} shows these $I_{ds}$ curves in the saturation region to the right of dashed line, generated via Spice, where
$V_{gs}$ is varied in the discrete set, $0.2,0.3,...,1~\rm{V}$ ($28~\rm{nm}$ Silicon technology model MOSFET is used for illustration purpose).
We can notice that the slope of the current curves increases as $V_{gs}$ increases due to CLM, which we leverage to perform the decoding at the receiver, as explained below.
$I_{ds}$ encodes the values of $V_{gs}$ ($x_2$) and $V_{ds}$ ($x_1$)---as opposed to extracting the length of the curve from origin to the mapped point, as in~\cite{Zhao2016, Zhao2018a}. 
It is necessary to have a discrete set of y-axis ($V_{gs}$) values, and the actual y-axis value is mapped to the nearest value from the set and applied to the MOSFET to generate the encoded current (Fig.~\ref{fig:ajscc_mos_3d}). The amount of the quantization between the y-axis values is $\phi$.
While the above MOSFET-based space-filling technique satisfies ($i$) and ($ii$) properties mentioned above, it violates ($iii$) as a given $I_{ds}$ value
could be generated from multiple pairs of $V_{gs}$ and $V_{ds}$ values (Fig.~\ref{fig:ajscc_mos_3d}).
This is problematic as it is difficult to decode the correct $V_{gs}$ at the receiver. To address this challenge, we design a decoding technique at the digital CHs based on the previously received $I_{ds}$ value in Sect.~\ref{sec:prop_soln:kldiv}.
We also note that our analog sensors would employ the Frequency Position Modulation and Multiplexing~(FPMM)~\cite{Sadhu2018ucomms}
to communicate with the digital CHs. FPPM consumes low power (of the order of few $\mu W$ theoretically) due to a very low SNR operating region at the receiver (about $-40~\rm{dB}$).

\subsection{Decoding and Estimation at Digital CHs (Receiver)}\label{sec:prop_soln:kldiv}
In this subsection, which deals with the digital nodes/CHs (receiver), we first describe how we perform the AJSCC decoding by solving the non-unique mapping problem described above; then, we present an approach using KDE and KLD to estimate the source distribution before describing how we optimize the AJSCC parameters based on the estimated distribution. Finally, we describe the analog circuit that performs MOSFET-based encoding for different $\phi$.

\textbf{(1)~Decoding at Receiver:}
We assume that the discrete set of $V_{gs}$ values 
used at the transmitter for encoding is known at the receiver. This is a valid assumption as the receiver decides the optimum $\phi$ to be used by the transmitter. The decoding process relies on the assumption that physical values do not change abruptly and hence two consecutive received $I_{ds}$ values at the receiver will lie on the same $I_{ds}$ curve (i.e., corresponding to a particular $V_{gs}$ value). For more details on the \textit{slope-matching technique} based decoding at receiver, refer~\cite{Sadhu2019iscas, Sadhu2018ucomms}.

\textbf{(2)~Estimating Source Distribution:}
Having knowledge of the source (transmitter) distribution at the receiver will help in determining the optimal AJSCC parameter, $\phi$. To illustrate this, consider the following scenarios where  adaptive tuning of parameters is necessary.
\emph{Fixed~region,~varying~accuracy: }As a hypothetical example, consider temperature data for climate monitoring in New Jersey. This will likely be in the range of -$10$ to $35$ Celsius degrees, with extreme temperatures less likely to be recorded. It is then natural to desire that the sensing accuracy within this range be high, while the accuracy out of it be low. Also, within this range, temperature readings will have a certain distribution; hence, it is desirable to have a varying accuracy (and, hence, variable parameters) over the entire range. %
\emph{Varying~regions,~varying~accuracy: }This corresponds to situations in which the monitoring application may require different accuracies for different deployment regions (e.g., consider an application where the sensing depends on the amount of solar radiation; as the Sun moves across the sky, some regions may be irradiated more than others). 
In order to estimate the source distribution at the receive based on the received values, we design an approach based on KDE and KLD as follows.

We now present an approach to estimate the distribution of $x_1$ at the source/transmitter based on their decoded values at the receiver---$\hat{x_1}$. The same procedure can be similarly applied for $x_2$. Let us denote the decoded values at the receiver as $y_1,...,y_k,...,y_K$. Kernel density estimator for these values can be written as,
$p_Y (y) = \frac{1}{{Kh}}\sum\limits_{k = 1}^K {f\left( {\frac{{y - y_k }}{h}} \right)}$,
where $p_Y (y)$ is the estimated density at the receiver, $K$ is the number of samples in the received data, $h$ is the kernel bandwidth, and $f()$ is the type of kernel function which can be chosen to be normal, uniform, rectangular, etc.
Assume the density at source is $p_X (x)$, then the KLD between densities $p_X (x)$, $p_Y (y)$ is expressed as,
$D_{KL} (X||Y) = \int\limits_{ - \infty }^\infty  {p_X (x)\log \frac{{p_X (x)}}{{p_Y (y)}}} dx$
Given the above formulations, the goal is to estimate $p_Y (y)$ that is closest to $p_X (x)$. For this purpose, we assume the source distribution is drawn from a set of known distributions. The receiver's objective is to estimate the exact source distribution out of the set of possible source distributions. This can be expressed as an optimization problem where the objective is to find the entities, $h_{opt}, f_{opt}$ that minimize the K-L divergence between $p_X (x)$ and $p_Y (y)$, where $p_X (x)$ is drawn from a set of known distributions, as expressed below,
\begin{equation*}
\begin{array}{l}
 \{ h_{opt} ,f_{opt} ( \cdot )\}  = \mathop {\min }\limits_{h,f( \cdot )} \left\{ {D_{KL} (p_X (x), p_Y (y))} \right\} \\
  =\mathop {\min }\limits_{h,f( \cdot )} \left\{ {\int\limits_{ - \infty }^\infty  {p_X (x)\log \frac{{p_X (x)}}{{p_Y (y)}}} dx} \right\} \\
  =\mathop {\min }\limits_{h,f( \cdot )} \left\{ {\int\limits_{ - \infty }^\infty  {p_X (x)\left\{ {\log p_X (x) - \log \left[ {\frac{1}{{Kh}}\sum\limits_{k=1}^K {f\left( {\frac{{y - y_k }}{h}} \right)} } \right]} \right\}} dx} \right\} \\
   =\mathop {\min }\limits_{h,f(\cdot)} \left\{{-\int\limits_{-\infty }^\infty  {p_X (x)\left\{ {\log \left[ {\frac{1}{{Kh}}\sum\limits_{k=1}^K {f\left( {\frac{{y - y_k }}{h}} \right)} } \right]} \right\}} dx}\right\}.
 \end{array}
 \end{equation*}
This optimization problem is non-linear and the solution can be found by searching the parameter space.
\begin{figure}
\begin{center}
\includegraphics[width=2.1in,height=2.2in]{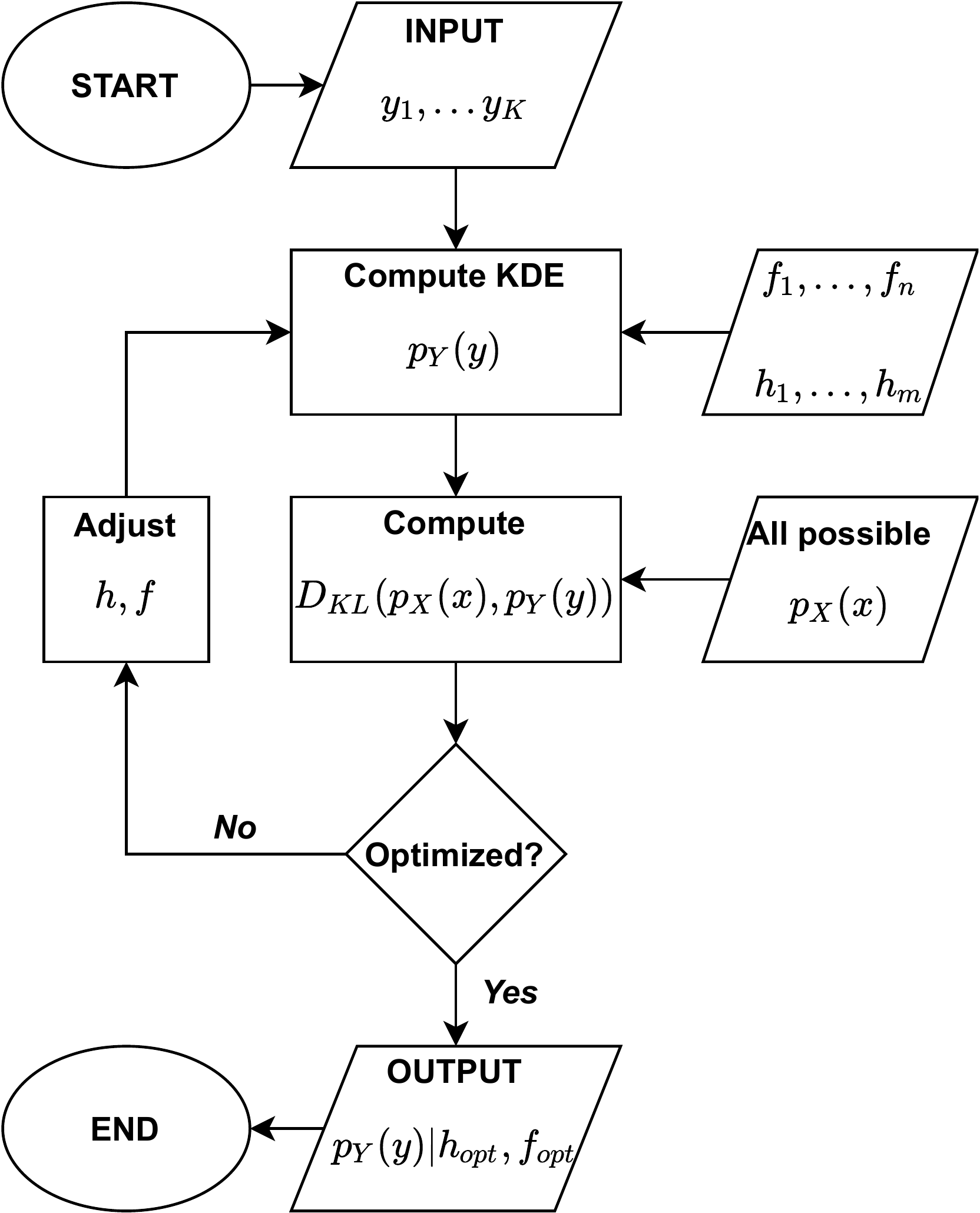}
\end{center}
\vspace{-0.15in}
\caption{Algorithm flowchart of the proposed iterative Kullback-Leibler Divergence~(KLD)-based Kernel Density Estimation~(KDE) algorithm to estimate the source distribution at the receiver.
}\label{fig:kde_kldiv}
\vspace{-0.2in}
\end{figure}
Fig.~\ref{fig:kde_kldiv} shows the procedure followed at the receiver as a flowchart to estimate the source distribution. The optimization process searches the parameter space of bandwidth, $h$ ($h_1 < h < h_m$) and kernel function $f()$ ($f \in \{f_1,...,f_n\})$, to minimize the K-L divergence score. The receiver uses a set of all possible source distributions for $p_X(x)$ to compute the KLD and choose the one with the minimum score.

Once the source distribution is estimated, the receiver can optimize the AJSCC configuration parameters such as $\phi$ as explained later below. The sensors can then be designed to also receive such simple signaling/configuration information from the digital nodes from two perspectives---(i)~the MOSFET-based encoding discussed in Sect.~\ref{sec:prop_soln:ajscc} need to be modified to accept the quantization parameter, $\phi$; (ii)~a duplex system need to be adopted for sensors' receiver design that uses different bands for transmitting and receiving so that the analog sensors can continuously transmit data and also receive configuration information periodically from the digital CHs. In order to minimize the power consumption in this setting, the duration for which the sensors receive data can be minimized by having a very low, possibly adaptive, duty cycle, say $\approx 5\%$. While (ii) is outside the scope of this article, we provide a variable $\phi$ circuit design to address (i), later below.

\textbf{(3)~AJSCC Parameter Optimization:}
The receiver decides the optimum $\phi$ based on a tradeoff among different quantities including spatial and temporal correlation of the phenomenon being sensed, of the communication channel, power budget available at the sensors, and minimum Mean Square Error~(MSE) requirements of the application. 
A smaller $\phi$ results in better MSE at the receiver as the quantization error is minimized; however, it also results in higher power consumption at the sensor (due to additional hardware) as higher resolution is sought.
As mentioned earlier, a smaller $\phi$ is more susceptible to channel noise as the decoded point may lie on a different $I_{ds}$ curve and vice-versa. Let us quantify the spatial and temporal correlation of the phenomenon being sensed using two variables---$s_p$ and $t_p$, respectively---where the former indicates the range/radius within which the phenomenon has spatial correlation/similarity while the latter indicates the time interval during which the phenomenon has temporal correlation/similarity. We assume the receiver estimates these parameters and hence, is aware of them.
These parameters can be leveraged at the receiver to increase $\phi$ (i.e., increase the quantization error), which can be compensated by averaging ($\overline{MSE}$) within both time $t_p$ and space $s_p$ (as values do not change significantly within these time windows/space ranges),
and thereby save on power. The wireless channel also plays a role in this process. Let us quantify the spatial and temporal correlation of the channel as $s_c$ and $t_c$, respectively, similarly to $s_p$ and $t_p$ defined above. Now, if $t_c < t_p$, \emph{time diversity} can help improve the MSE at the receiver; similarly, if $s_c < s_p$, \emph{space diversity} can help. If time and/or space diversity exist, it is possible to have good average $\overline{MSE}$ at the receiver. However, when both are absent, i.e., $t_c > t_p$ \emph{and} $s_c > s_p$ (i.e., neither time nor space diversity can be exploited), the effective channel condition determines the MSE at the receiver.

\textbf{(4)~Variable $\phi$ Circuit Design:}
As motivated previously, there are scenarios in which there is a need to change the $\phi$ \emph{adaptively}.
Hence, we develop a circuit design for MOSFET-based AJSCC encoding that accepts different levels of $\phi$, specifically, we design for $\phi=1, 0.5, 0.25, 0.125~\rm{V}$. For more details on the design of this circuit, refer~\cite{Sadhu2019iscas}.

\section{Performance Evaluation}\label{sec:perf_eval}
We provide results on the analog sensing (Sect.~\ref{sec:perf_eval:ajscc}) and on the digital receiver estimation/optimization (Sect.~\ref{sec:perf_eval:kldiv}), followed by a summary and discussion (Sect.~\ref{sec:perf_eval:dis}).

\begin{figure*}[ht]
        \centering
        \hspace{-0.25in}
        \begin{subfigure}[b]{0.32\textwidth}
        		\centering
        	 \includegraphics[width=1\textwidth,height=1.7in]{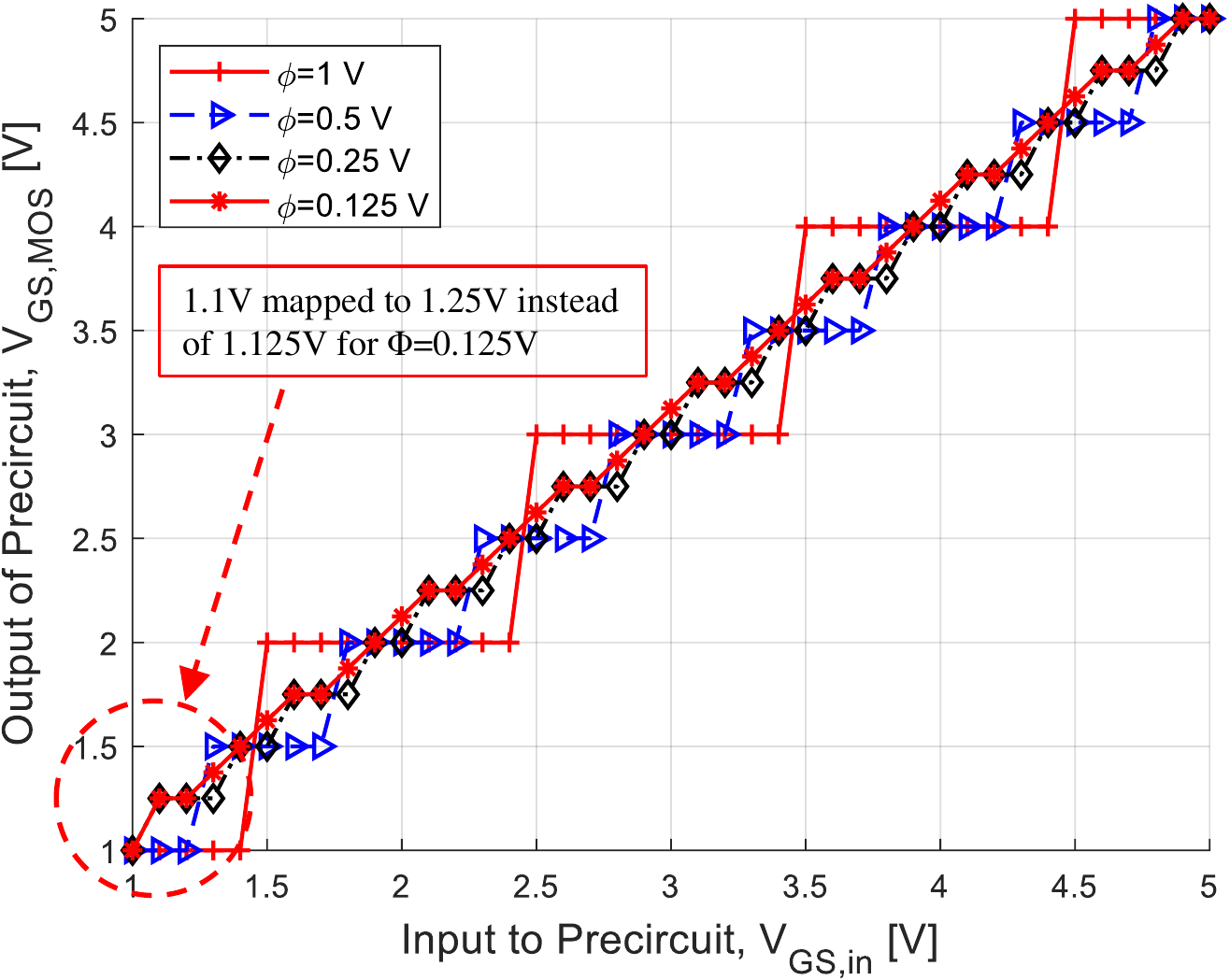}
        		\caption{}
        		\label{fig:precircuit_result}
        	\end{subfigure}
        	 \hspace{-0.15in}
~        	 
           \begin{subfigure}[b]{0.32\textwidth}
        		\centering
        	 \includegraphics[width=1\textwidth,height=1.7in]{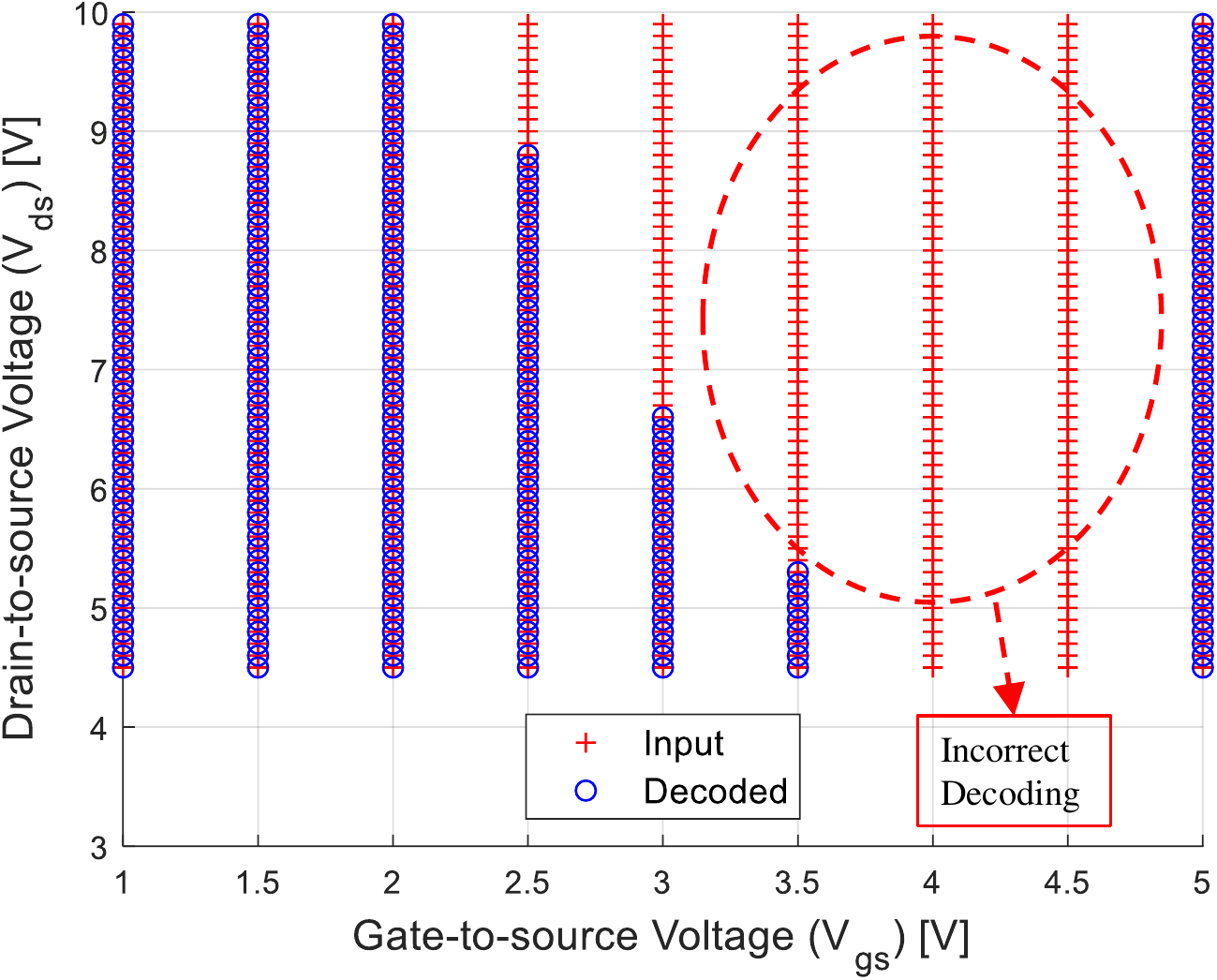}
        		\caption{}
        		\label{fig:before_correction_delta_05}
        	\end{subfigure}
        	 \hspace{-0.15in}
~        	 
        \begin{subfigure}[b]{0.32\textwidth}
            \centering
            \includegraphics[width=1\textwidth,height=1.7in]{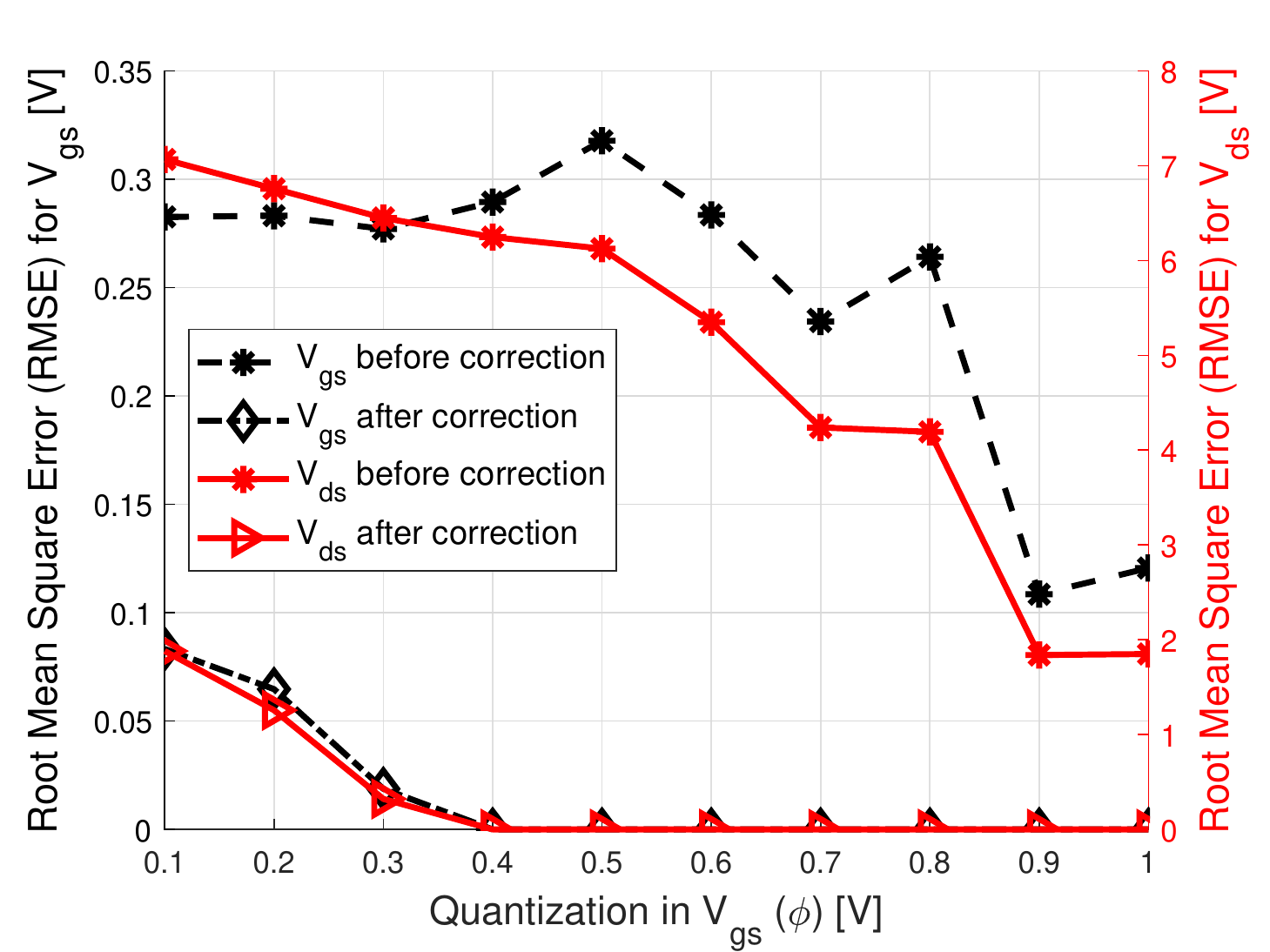}
            \caption{}
            \label{fig:rmse_phi}
        \end{subfigure}
        \vspace{-0.1in}
        \caption{\label{fig:precircuit_enc_dec}(a)~$V_{gs,in}$ mapped to $V_{gs,MOS}$ by the precircuit for different $\phi$ values; (b)~Decoding results for $\phi=0.5~\rm{V}$ when no correction logic is used; (c)~Root Mean Square Error~(RMSE) of $V_{gs}$ and $V_{ds}$ before and after correction logic is applied as $\phi$ is varied.}
        \vspace{-0.2in}
\end{figure*}

\begin{figure*}[ht]
        \centering
        \hspace{-0.25in}
        \begin{subfigure}[b]{0.33\textwidth}
        		\centering
        	 \includegraphics[width=1.1\textwidth,height=1.7in]{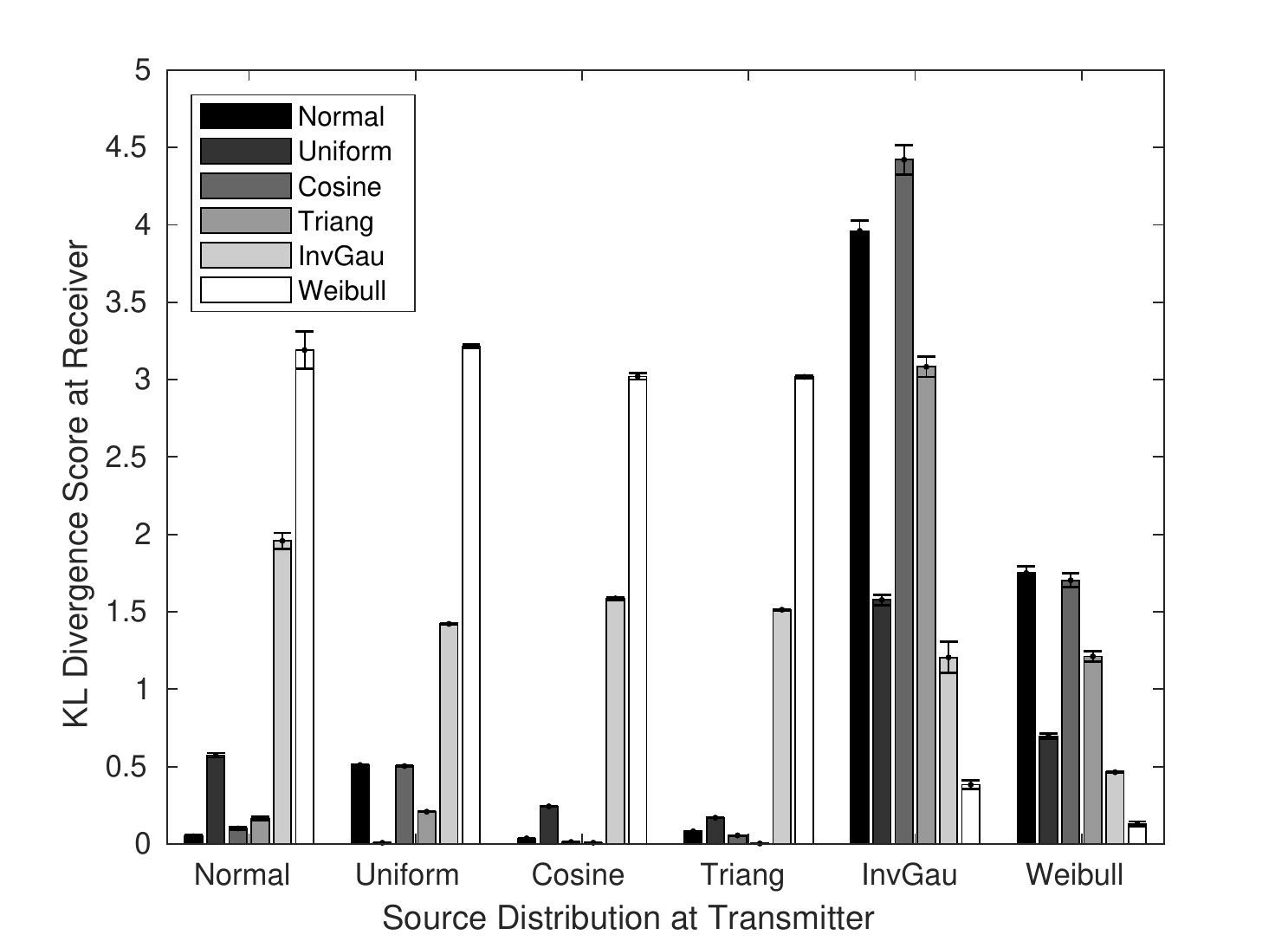}
        		\caption{}
        		\label{fig:kldivs_vs_dists_var_dists_x1}
        	\end{subfigure}
        	 \hspace{-0.15in}
~        	 
           \begin{subfigure}[b]{0.33\textwidth}
        		\centering
        	 \includegraphics[width=1.1\textwidth,height=1.7in]{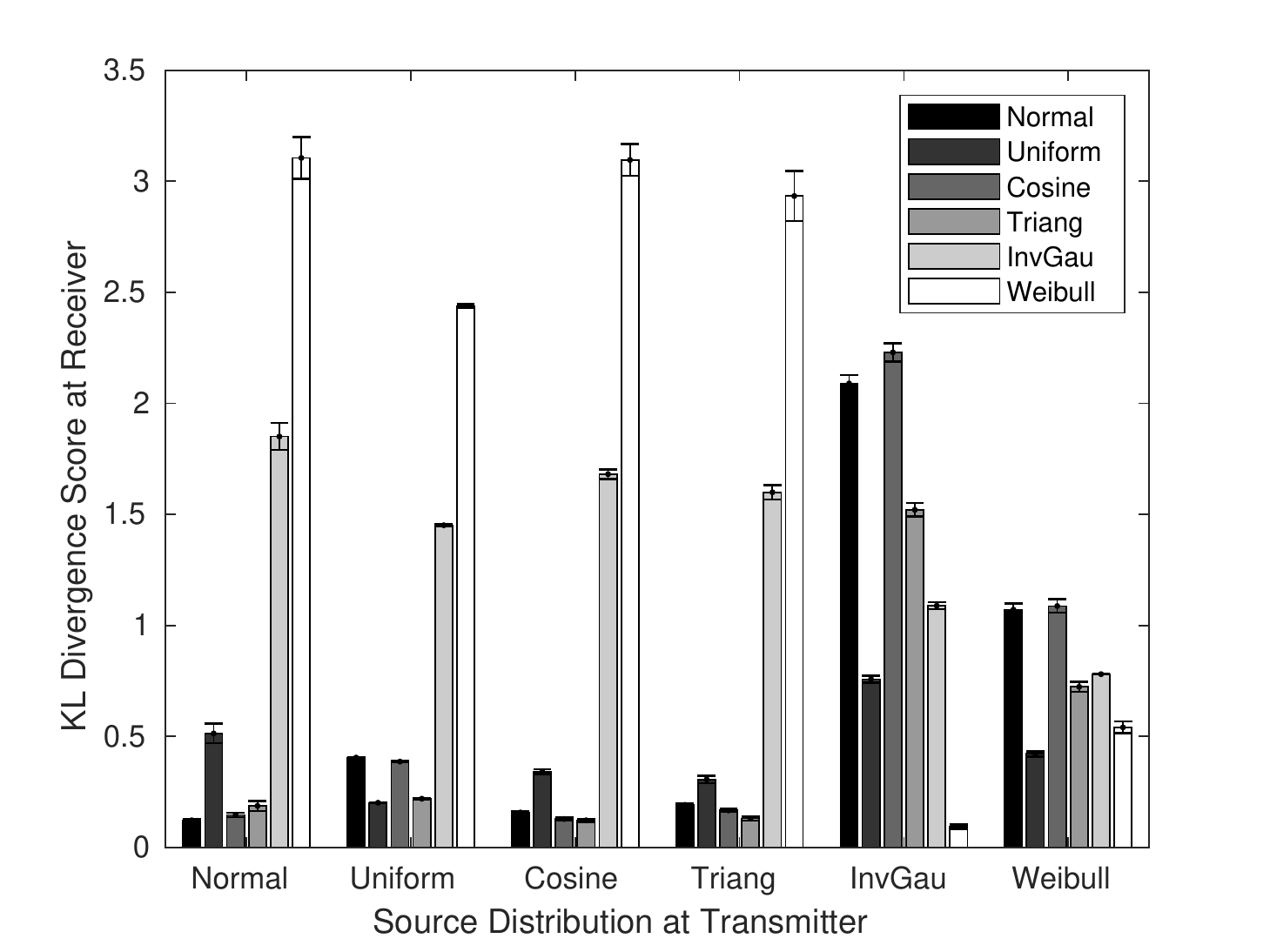}
        		\caption{}
        		\label{fig:kldivs_vs_dists_var_dists_x2}
        	\end{subfigure}
        	 \hspace{-0.15in}
~        	 
        \begin{subfigure}[b]{0.33\textwidth}
            \centering
            \includegraphics[width=1.1\textwidth,height=1.7in]{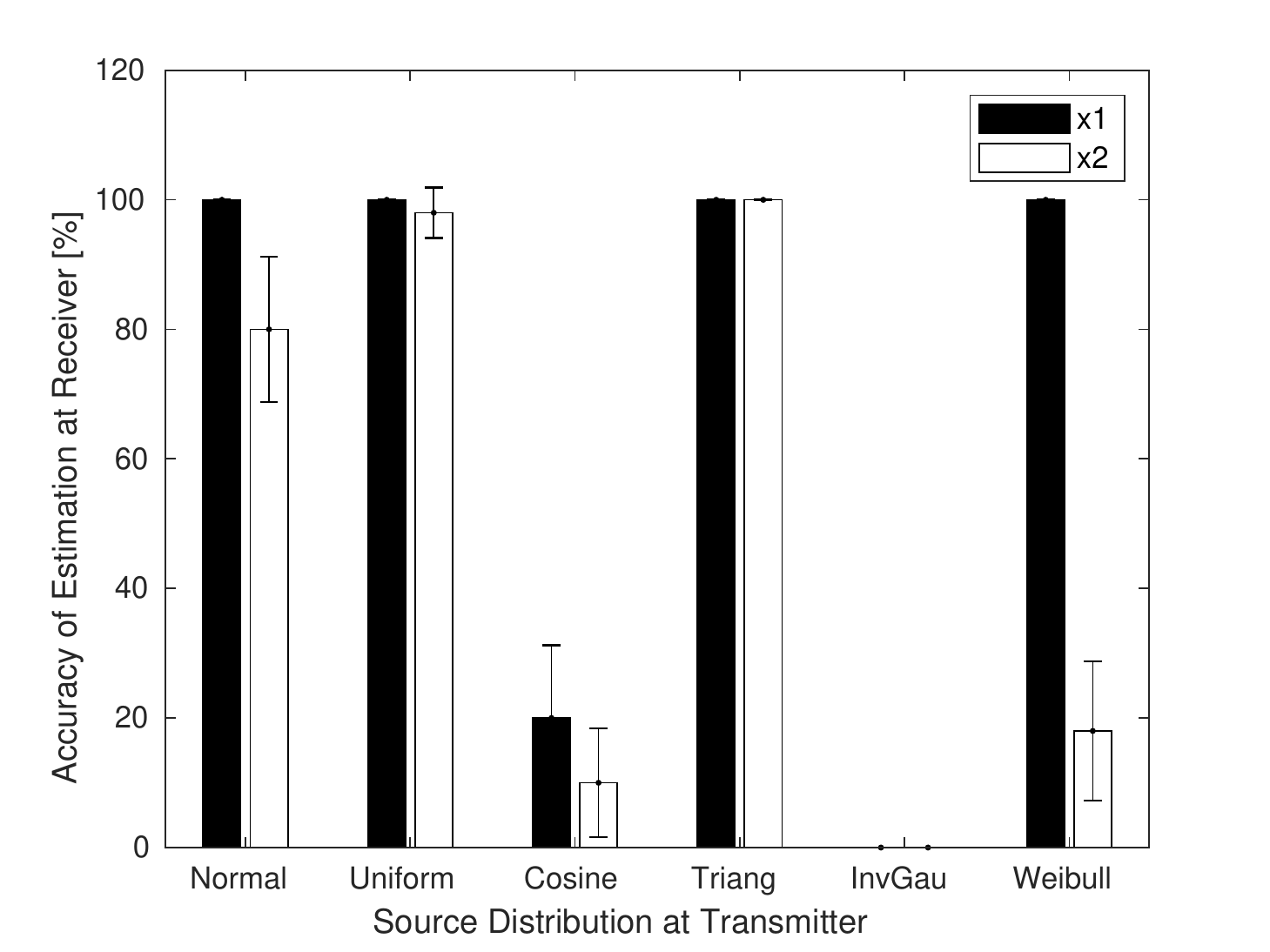}
            \caption{}
            \label{fig:accu_vs_dist_var_x1,x2}
        \end{subfigure}
        \vspace{-0.1in}
        \caption{\label{fig:results1a}KLD score values calculated at the receiver by comparing the estimated KDE distribution of the received values with each of the known source distributions as the source distributions are varied for (a)~$x_1$ ($V_{ds})$; (b)~$x_2$ ($V_{gs}$); (c)~Accuracy for $x_1$ and $x_2$ choosing the source distribution with the lowest KLD score as the estimated source distribution. The parameters are fixed to $\phi=0.2$, SNR=$20~\rm{dB}$ and bandwidth=$200~\rm{KHz}$.}
        \vspace{-0.2in}
\end{figure*}

\subsection{Analog Sensing Substrate}\label{sec:perf_eval:ajscc}
In this subsection, we evaluate the functionality of the proposed MOSFET-based encoding and decoding (including precircuit design) using MATLAB and LTSpice simulations.

\textbf{Precircuit:}
To verify the functionality of the precircuit, we varied $V_{gs,in}$ from $1$ to $5~\rm{V}$, in increments of $0.1~\rm{V}$, for all four $\phi$ values. The reason not to start from $0~\rm{V}$ is that $V_{gs}$ should be greater than the threshold voltage, $V_{th} \approx 0.8~\rm{V}$, for the MOSFET to operate. The results, shown in Fig.~\ref{fig:precircuit_result}, are as expected for the case of $\phi=1,0.5~\rm{V}$. However, for the case of $\phi=0.25,0.125~\rm{V}$, the circuit maps to one level higher than expected for some voltages. For example, when $\phi=0.125~\rm{V}$, $V_{gs,in}=1.1~\rm{V}$ is mapped to $1.25~\rm{V}$ instead of $1.125~\rm{V}$. The reason for this may be the saturation effect of the Operational Amplifier~(OpAmp) used in the adder. Curated circuit design optimizations, outside the scope of this work, can help circumvent this limitation. 

\textbf{Encoding and Decoding:}
We used a $0.18~\mu\rm{m}$ technology n-channel MOSFET~(nMOS) with $W\cdot\mu\cdot C_{ox}/L = 155\times 10^{-6}~\rm{F/Vs}$, $V_{th} = 0.74~\rm{V}$, $\lambda = 0.037~\rm{V^{-1}}$ for evaluation purposes. $V_{ds}$ is varied from $4.5$ to $10~\rm{V}$, in increments of $0.1~\rm{V}$. The reason not to start from $0~\rm{V}$ is to ensure that the MOSFET is well into the saturation region. Discrete set of $V_{gs}$ values in the range $[1,5]~\rm{V}$ as per $\phi$ are considered, e.g., $V_{gs}=1,2,3,4,5~\rm{V}$ for $\phi=1~\rm{V}$; hence, for each $V_{gs}$, 55 values of $V_{ds}$ are considered. Upon applying these voltages to the MOSFET, the generated $I_{ds}$ values are recorded and sent to the digital receiver (no wireless channel), where the decoding process is done. At the receiver, each curve is processed independently and two consecutive $I_{ds}$ values from the same curve are used for decoding the correct $V_{gs}$ using the slope-matching technique. The results are shown in Fig.~\ref{fig:before_correction_delta_05} for $\phi=0.5~\rm{V}$, where the original values are shown using `+' and decoded values using `o'. We can see that some of the values are decoded incorrectly (where there are bare `+' without `o'). 
The reason is due to the mismatch between two slopes---the slope calculated theoretically, $\lambda I_{ds}$ (which varies with $V_{ds}$), is an approximation (i.e., valid only for $\lambda V_{ds} << 1$) of the actual slope calculated using the two-point formula (which is independent of $V_{ds}$).
To solve this problem, we used a range-checking technique where, if the decoded $V_{ds}$ value corresponding to the best (in terms of slope match) $V_{gs}$ value does not fall within the $V_{ds}$ range assumed at the transmitter $(4.5,10)~\rm{V}$, the next best $V_{gs}$ value (in terms of slope match) is chosen, and the process is repeated iteratively. %
Using this correction logic, we are able to improve decoding accuracy.
To see the effect of $\phi$ on the decoding process, we varied it from $0.1$ to $1~\rm{V}$. Figure~\ref{fig:rmse_phi} shows the Root Mean Square Error~(RMSE) in $V_{gs}$ and $V_{ds}$ before and after the correction logic is applied. We have used separate axes for $V_{gs}$ and $V_{ds}$ as the error is higher in the case of $V_{ds}$. We can notice the following---(i)~before correction, the errors are high, up to about $0.3~\rm{V}$ for $V_{gs}$ and $7~\rm{V}$ for $V_{ds}$; (ii)~after correction, the error reduces up to about $0.1~\rm{V}$ for $V_{gs}$ and $2~\rm{V}$ for $V_{ds}$; (iii)~RMSE $\approx 0$ for $\phi \geq 0.4~\rm{V}$; increases steadily for $< 0.4~\rm{V}$.

\textbf{Power Consumption:}
Our encoding design consists of precircuit and MOSFET. The power consumed by MOSFET is negligible compared to that of precircuit. Our precircuit primarily consists of OpAmps, comparators, multiplexers, and resistors, of which OpAmps are clearly the major contributors to the overall power consumption. The precircuit consumes one OpAmp for each stage and one for the final adder. For comparison purposes and to get an estimate of power consumption when our circuit is fabricated using the latest $nm$-Silicon technology, we use the same low power nano designs for the above components as considered in~\cite{Zhao2016} ($8~\mu\rm{W}$ for OpAmp and $12.7~\rm{nW}$ for comparator). For 9 AJSCC levels ($\phi=0.5~\rm{V}$, 2-stages), the power consumption is $\approx 24~\mu\rm{W}$. On the other hand, Design~1~\cite{Zhao2016} with $11$ levels consumes $130~\mu\rm{W}$ and Design~2~\cite{Zhao2018a} with $8$ levels consumes $64~\mu\rm{W}$.
\begin{figure*}[ht]
        \centering
        \hspace{-0.25in}
        \begin{subfigure}[b]{0.33\textwidth}
        		\centering
        	 \includegraphics[width=1.1\textwidth,height=1.7in]{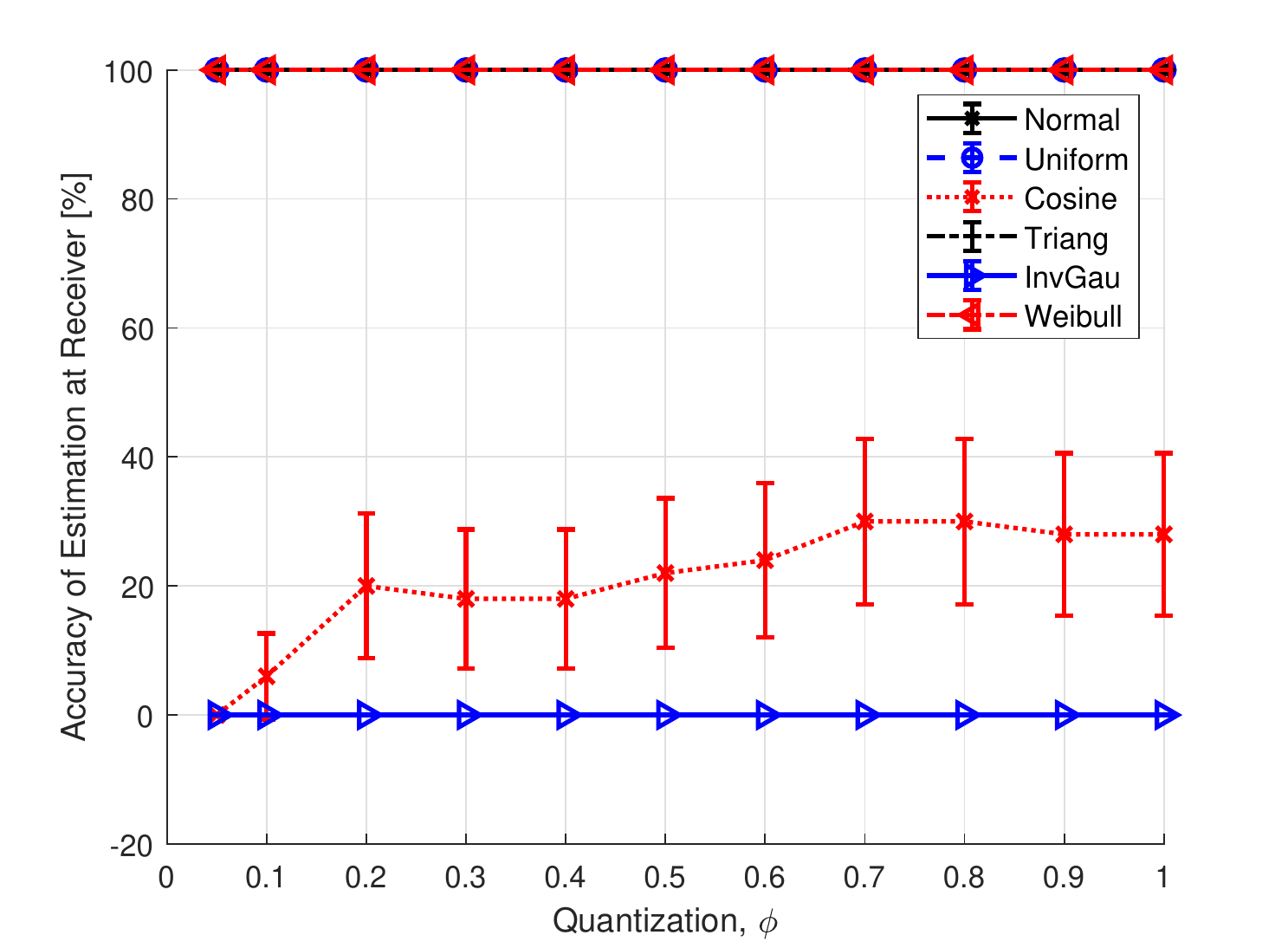}
        		\caption{}
        		\label{fig:accu_vs_delta_var_dist_x1}
        	\end{subfigure}
        	 \hspace{-0.15in}
~        	 
           \begin{subfigure}[b]{0.33\textwidth}
        		\centering
        	 \includegraphics[width=1.1\textwidth,height=1.7in]{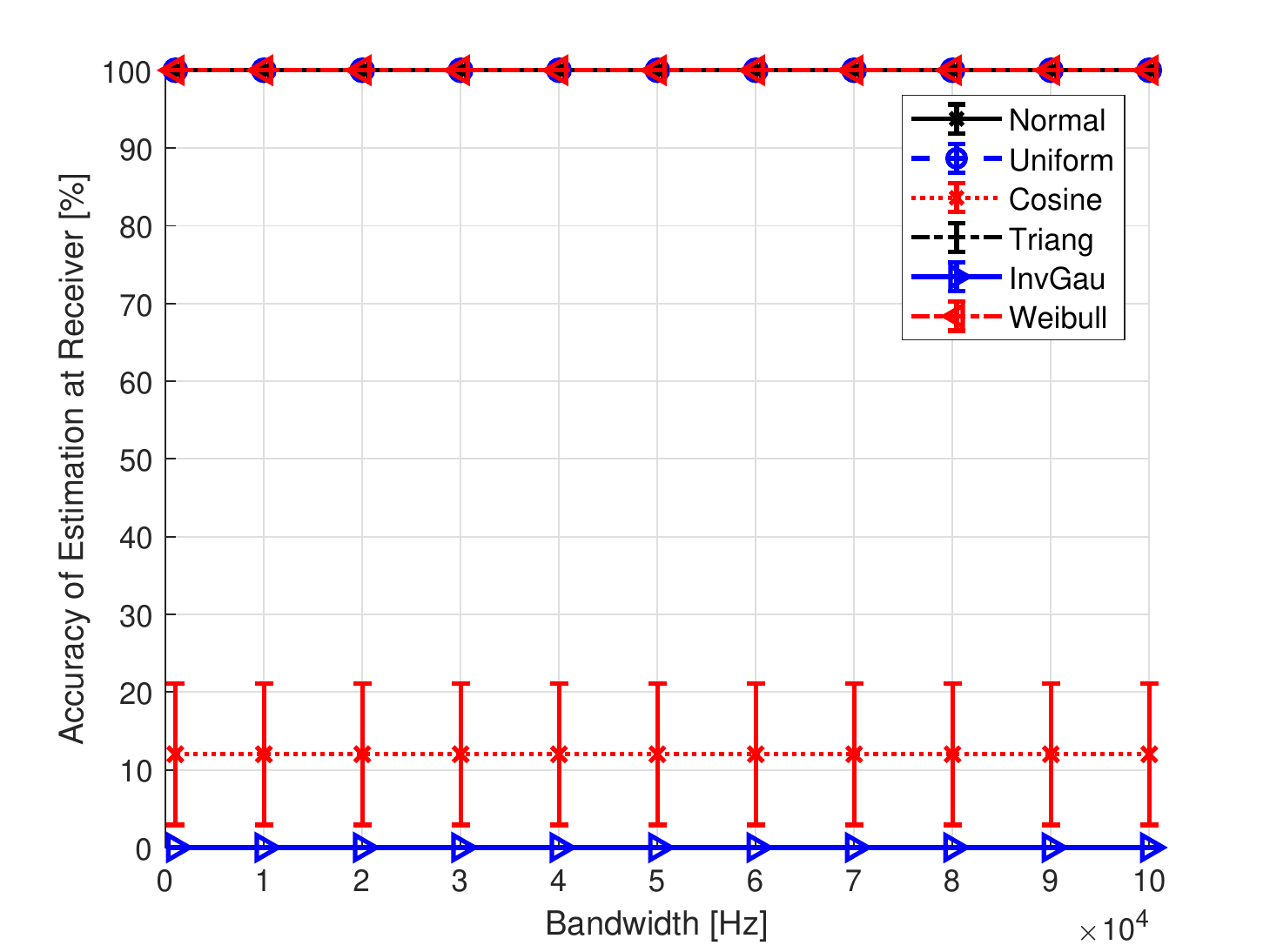}
        		\caption{}
        		\label{fig:accu_vs_bw_var_dist_x1}
        	\end{subfigure}
        	 \hspace{-0.15in}
~        	 
        \begin{subfigure}[b]{0.33\textwidth}
            \centering
            \includegraphics[width=1.1\textwidth,height=1.7in]{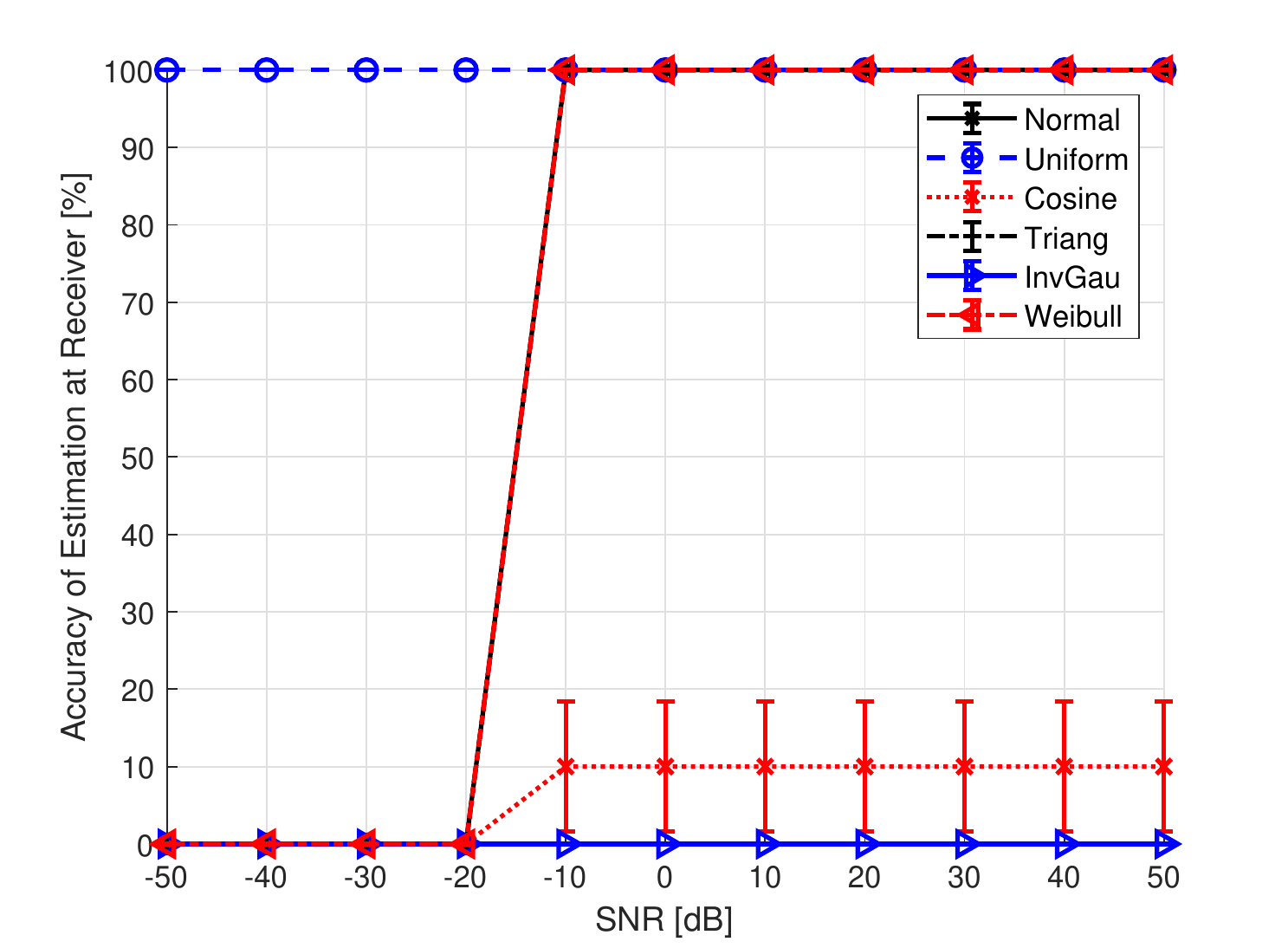}
            \caption{}
            \label{fig:accu_vs_snr_var_dist_x1}
        \end{subfigure}
        \hspace{-0.15in}
~        
        \begin{subfigure}[b]{0.33\textwidth}
        		\centering
        	 \includegraphics[width=1.1\textwidth,height=1.7in]{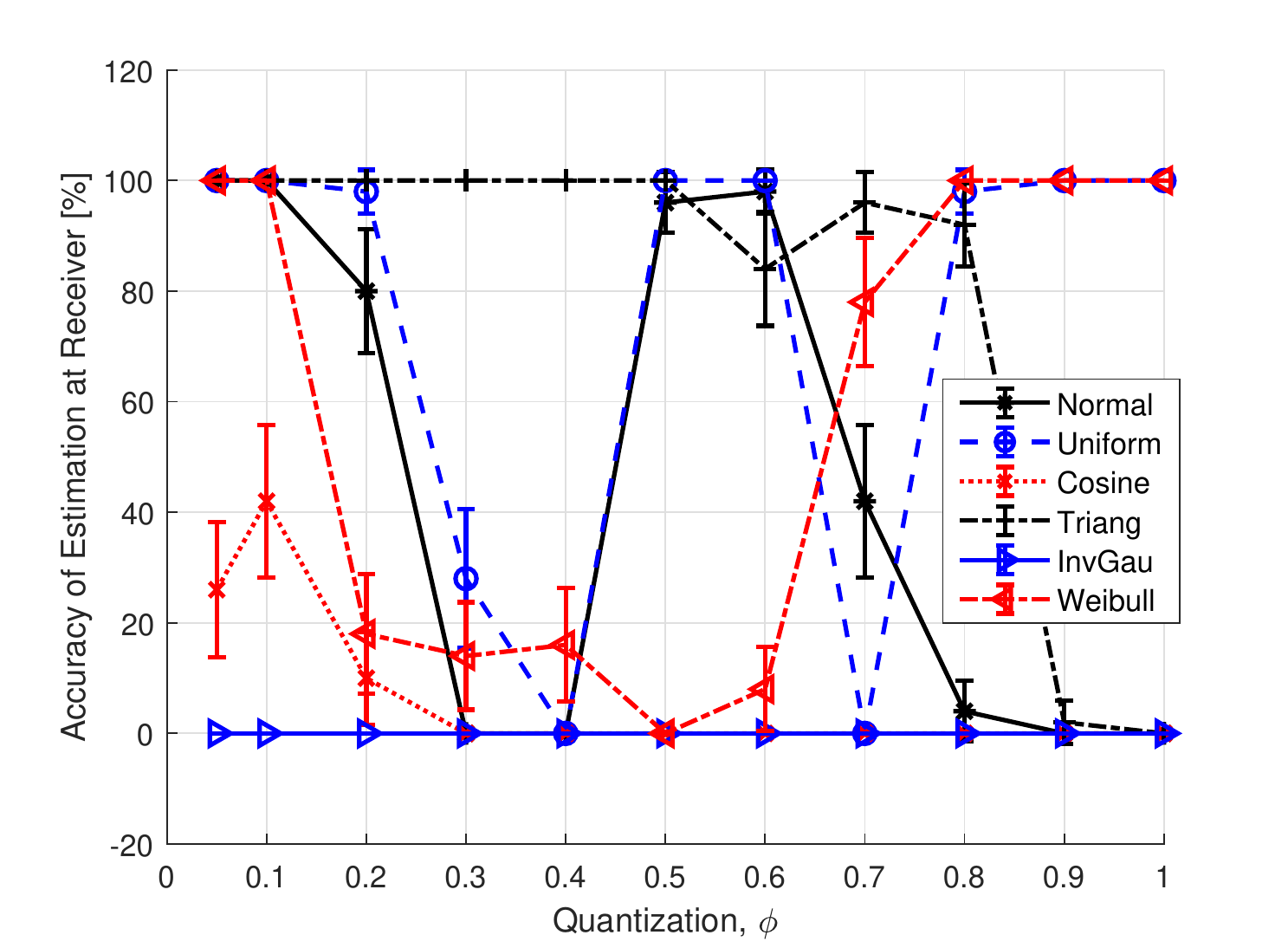}
        		\caption{}
        		\label{fig:accu_vs_delta_var_dist_x2}
        	\end{subfigure}
        	 \hspace{-0.15in}
~        	 
       \begin{subfigure}[b]{0.33\textwidth}
        		\centering
        	 \includegraphics[width=1.1\textwidth,height=1.7in]{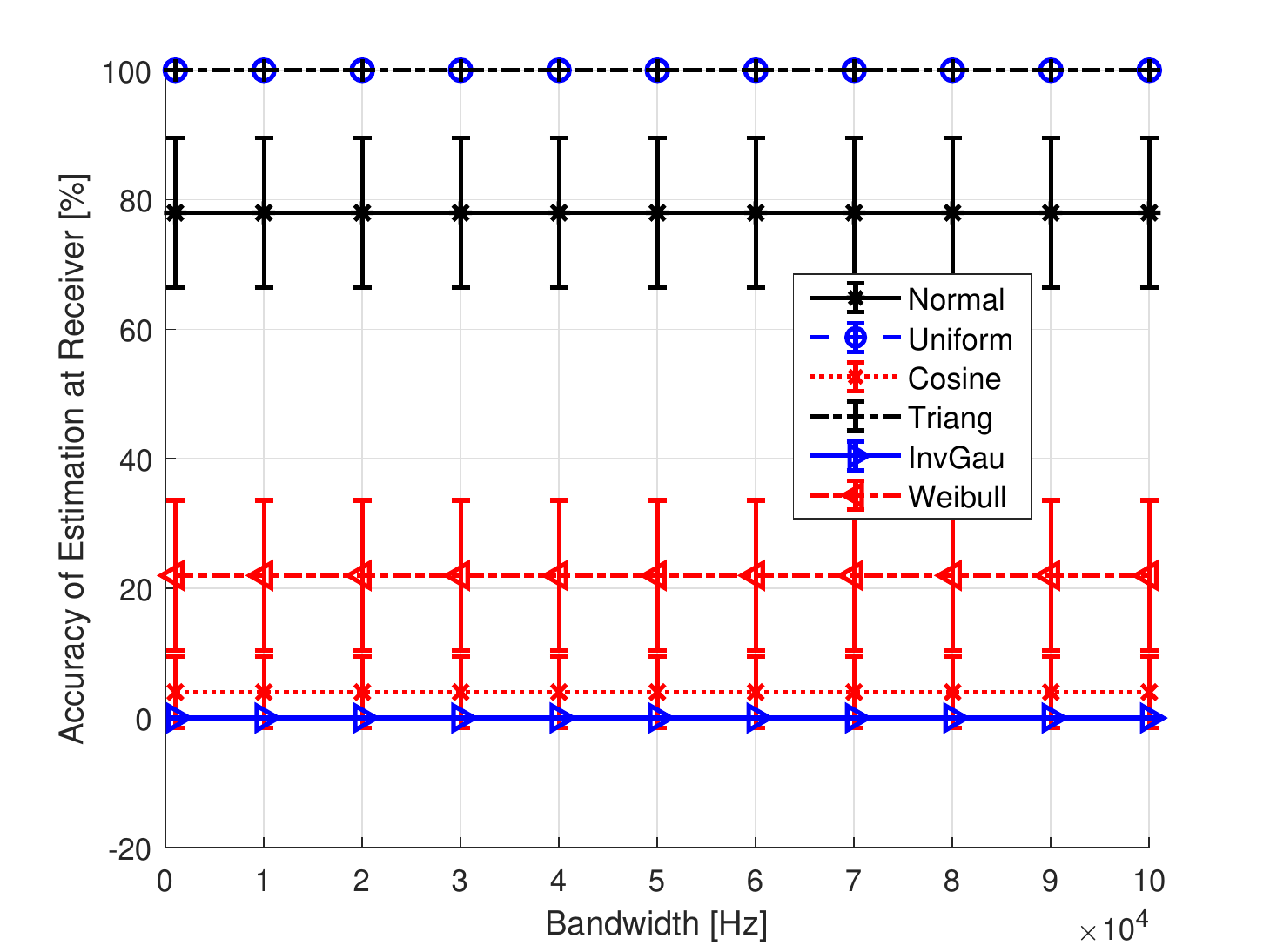}
        		\caption{}
        		\label{fig:accu_vs_bw_var_dist_x2}
        	\end{subfigure}
        	 \hspace{-0.15in}
~        	 
        \begin{subfigure}[b]{0.33\textwidth}
            \centering
            \includegraphics[width=1.1\textwidth,height=1.7in]{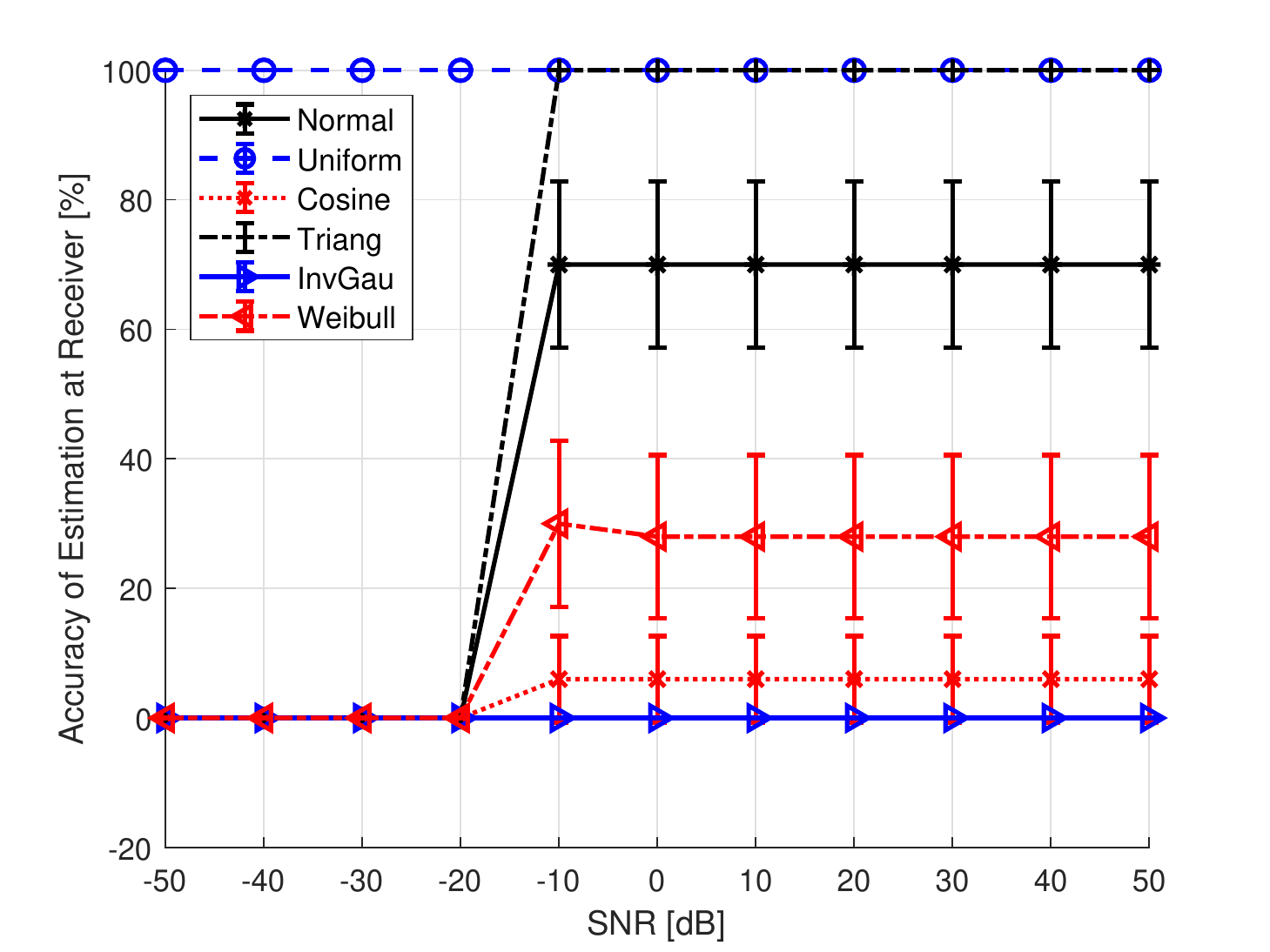}
            \caption{}
            \label{fig:accu_vs_snr_var_dist_x2}
        \end{subfigure}
        \vspace{-0.1in}
        \caption{\label{fig:results1b}Accuracy of the source distribution estimation at the receiver for different source distributions as the following design and channel parameters are varied: $\phi$, SNR, Bandwidth (BW), fixing other two parameters at $\phi=0.2$, SNR=$20~\rm{dB}$ and bandwidth=$200~\rm{KHz}$ for (a)-(c)~$x_1$ ($V_{ds})$; (d)-(f)~$x_2$ ($V_{gs})$.}
        \vspace{-0.2in}
\end{figure*}

\subsection{Estimation and Optimization at Digital CHs}\label{sec:perf_eval:kldiv}
In this subsection, we evaluate the proposed KDE and KLD based source estimation algorithm, followed by finding the optimal $\phi$ values based on source distribution and channel conditions.

\textbf{Source Estimation:}
To study the performance of the proposed KDE and KLD-based source estimation algorithm, we ran MATLAB simulations with the following setting. We considered a $20 \times 20$ array of sensors generating values for 20 time instants (hence, %
$20 \times 20 \times 20$ values). We considered $s_p = t_p = 10$. This means that, at any given time instant, $10 \times 10$ subarrays (four in our case) have spatially similar values and, for each sensor, for a duration of $10$ time instants, the values are temporally similar. These temporally and spatially similar values in the range $[0,1]$ are generated from one of the six distributions---\textit{`normal', `uniform', `cosine', `triangular', `inverse gaussian' (denoted `invgau' for simplicity) and `weibull'} with standard parameters 
(for details on these distributions, refer to~\cite{randraw}). We have selected these distributions as they represent a wide variety of commonly found distributions in nature and also considered some similar distributions so as to `confuse' and test the estimation system.

The receiver is aware that the source distribution belongs to one of these six distributions. Two such instances have been considered, one for $V_{gs}$ and one for $V_{ds}$. Since the values generated are between 0 and 1, they are scaled and offset to lie between $(5,10)~\rm{V}$. This is done to capture the saturation behavior (in case of $V_{ds}$) and to ensure that $V_{gs}$ values are far from $V_{th}$. The error in decoding is very high when $V_{gs}$ is close to $V_{th}$ as the term $(V_{gs} - V_{th})^2$ appears in the denominator of $\partial V_{ds}/\partial V_{gs}$; hence, it is desirable to range $V_{gs}$ values far away from $V_{th}$. The $V_{gs}$ values have been quantized in accordance with Shannon mapping, before generating the encoded values using~\eqref{eq:ids_clm}. 
The encoded $I_{ds}$ values have been frequency modulated (i.e., each value is mapped to a frequency using, for example, a scaling factor) and then passed through a Rician channel in MATLAB with single path, and a Doppler shift equal to $2\%$ of the transmitting frequency. Then, Additive White Gaussian Noise~(AWGN) noise as per bandwidth~(BW) and SNR considered is added. At the receiver, the values are first passed through Fast Fourier Transform~(FFT) analysis to identify the peak frequency value, which is then mapped back to find the $\hat{I}_{ds}$ value using frequency demodulation (i.e., using the same scaling factor as in the transmitter). The slope-matching technique is used to decode the respective $\hat{V}_{gs}$ and $\hat{V}_{ds}$ values.

Once the decoded values are found, the algorithm in Fig.~\ref{fig:kde_kldiv} is executed by considering the following parameters for KDE: four different kernels ($n=4$), viz., `normal', `box', `triangle', `epanechnikov' are considered. The bandwidth parameter, $h$ is varied among $m=10$ different values from $0.1,0.2,...,1$.  The obtained KDE is compared with all the six known distributions to find the KDE scores. The distribution which has the lowest KDE score is estimated to be the source distribution. Figs.~\ref{fig:kldivs_vs_dists_var_dists_x1},~\ref{fig:kldivs_vs_dists_var_dists_x2} show the KLD scores calculated at the receiver by comparing the estimated KDE with each of the six known source distributions. For this purpose, the parameters have been set to $\phi=0.2$, SNR=$20~\rm{dB}$ and bandwidth=$200~\rm{KHz}$. We can observe that the `cosine' and 'invgau' fare badly compared to others. This is because, the receiver confuses `cosine' as `triangular' and `invgau' as `weibull' as these distributions are similar. We can also note that $x_2$ performs poorly compared to $x_1$ due to the quantization error in the former. These two factors can be confirmed in Fig.~\ref{fig:accu_vs_dist_var_x1,x2} which shows the accuracy of the estimation by choosing the source distribution with the lowest score.

We then calculated the accuracy at the receiver by varying the design and channel parameters, $\phi$, SNR and bandwidth for different source distributions by fixing other two parameters at $\phi=0.2$, SNR=$20~\rm{dB}$ and bandwidth=$200~\rm{KHz}$. The results for $x_1$ are plotted in Figs.~\ref{fig:results1b}(a)-(b), while for $x_2$ in Fig.~\ref{fig:results1b}(d)-(f). For $x_1$, we can notice that `cosine' and `invgau' fare badly compared to others as explained before. However it is interesting to note the performance improvement for $SNR > -20~\rm{dB}$. For $x_2$, however, we note an interesting trend for variation with $\phi$. We observe that as $\phi$ is varied, the accuracy drops and then increases for different source distributions. The reason for this behavior is the quantization of $x_2$, which modifies the original distribution of $x_2$. In our case, for instance, while $x_2$ originally has cosine distribution, the quantized $\hat{x_2}$ at the receiver has normal distribution which is the cause for drop in accuracy. However this behavior tends to diminish as the quantization is very minute or is very large, while is pronounced for quantization values in between. From Fig.~\ref{fig:accu_vs_delta_var_dist_x2}, we can also infer which $\phi$ values perform well for each distribution, so the sensors can be programmed with those values accordingly.

\begin{figure}
\begin{center}
\includegraphics[width=3.7in,height=1.8in]{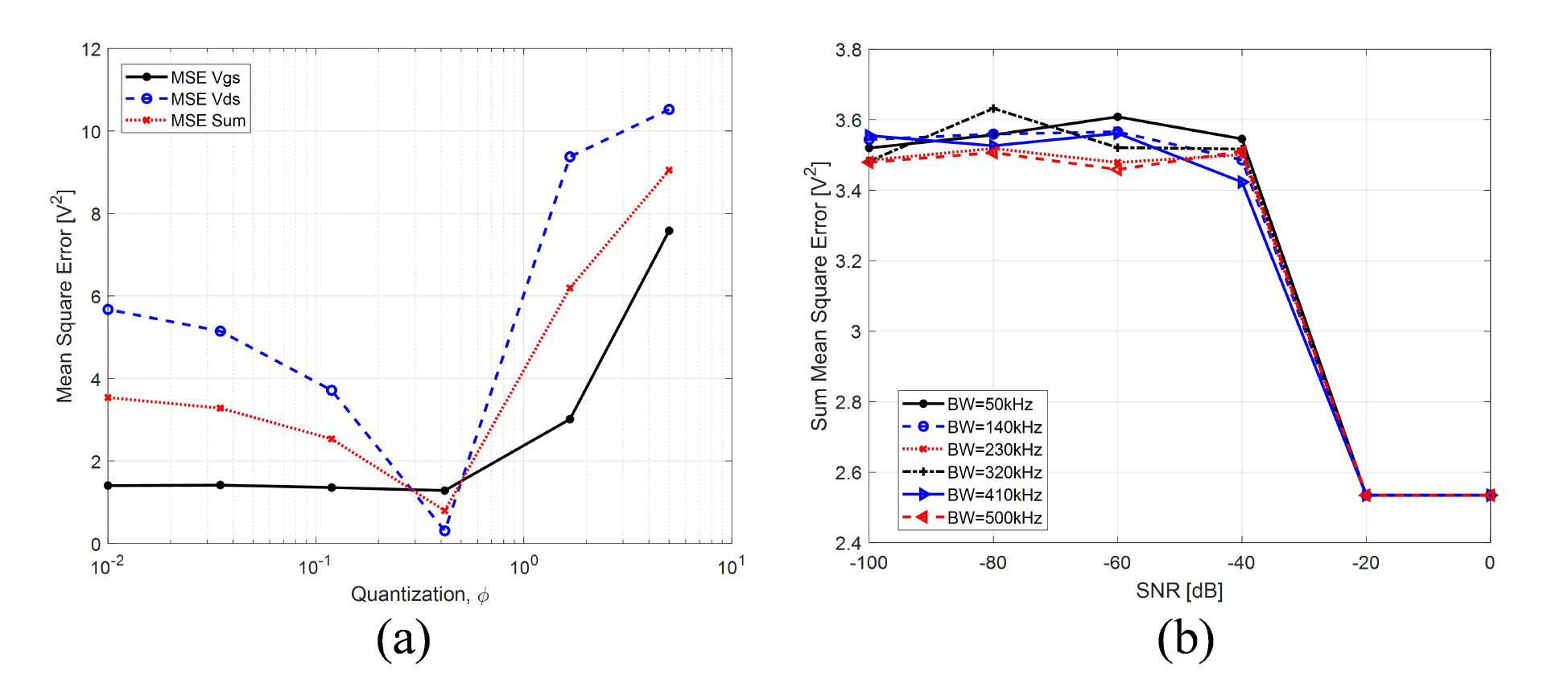}
\end{center}
\vspace{-0.2in}
\caption{(a)~Mean Square Error~(MSE) at receiver of $V_{gs}$, $V_{ds}$ and their sum vs. spacing between $I_{ds}$ curves ($\phi$) at transmitter (Bandwidth = $410~\rm{kHz}$, SNR = $-20~\rm{dB}$); (b)~MSE vs. SNR for different bandwidths ($\phi = 0.41$).}\label{fig:results_chan}
\vspace{-0.2in}
\end{figure}

\textbf{AJSCC Parameter Optimization:}
We study this optimization assuming the estimated source distribution at the receiver (from above) is uniform. Similar procedure can be followed in case the estimated distribution is non-uniform. We used the same simulation setup as above in order to study how MSE varies with channel conditions such as bandwidth~(BW) and SNR and find the optimal $\phi$ value.
After finding the decoded $\hat{V}_{gs}$ and $\hat{V}_{ds}$ values, $\overline{MSE}$ is found by averaging over both space ($s_p$) and time ($t_p$). Fig.~\ref{fig:results_chan}(a) shows the $\overline{MSE}$ of $V_{gs}$, $V_{ds}$ and their sum for BW=$410~\rm{kHz}$ and SNR=$-20~\rm{dB}$. We can notice that indeed an optimum $\phi^* = 0.41$ is achieved corresponding to $\overline{MSE}_{gs}=1.2~\rm{V^2}$ in $V_{gs}$, $\overline{MSE}_{ds} = 0.3~\rm{V^2}$ in $V_{ds}$ and $\overline{MSE}_{sum}=0.7~\rm{V^2}$ in their sum. Larger $\overline{MSE}_{ds}$ for smaller $\phi$ is attributed to the decoding process. For a very small $\phi$, it is possible that the decoded $\hat{V_{gs}}$ lies on adjacent levels to the actual one. However, since $\phi$ is small, it will result in minor MSE for $V_{gs}$ but not for $\hat{V_{ds}}$ because of~\eqref{eq:ids_clm}. Fig.~\ref{fig:results_chan}(b) shows the $\overline{MSE}_{sum}$ as the SNR is varied from $-100$ to $0~\rm{dB}$ for different bandwidths varying from $50$ to $500~\rm{kHz}$ with $\phi^*=0.41$ found from above. We can notice that for $SNR<-30~\rm{dB}$ there is a sharp decrease in performance. While the performance is approximately similar for all bandwidths considered, we can notice an improvement in SNR as the bandwidth is increased.

\subsection{Summary and Discussions}\label{sec:perf_eval:dis}

From the above results, we can infer the following main points: (i)~the precircuit works as expected, except for low values of $\phi \leq 0.25~\rm{V}$ when it quantizes to a level higher than expected occasionally; the precircuit consumes $\approx 24~\mu\rm{W}$ power; (ii)~the error in $V_{gs}$ and $V_{ds}$ at the receiver increases linearly when $\phi < 0.4~\rm{V}$ and is negligible for $\phi \geq 0.4~\rm{V}$; (iii)~the source distribution estimation at the receiver fares poorly---when input source distributions are very similar (e.g., `invgau' and `weibull'); estimation performance is poor in case of $x_2$ (than $x_1$) as the original distribution is modified due to quantization (e.g., `cosine' transforming to `uniform'); this behavior is less pronounced at medium (i.e., neither too high nor too low) quantization levels; (iv)~it is possible to find an optimal $\phi$ at the receiver considering the estimated source distribution, wireless channel conditions, etc., which can then be fed to the analog sensors.

The proposed solution quantizes one of the input signals and as such may not be suitable for all applications. Only applications that can tolerate a drop in accuracy in one of the parameters of interest---e.g., Humidity quantized to $5\%$---can benefit from our solution. Secondly, we are yet to test the performance of our solution with respect to the frequency characteristics of the sensing phenomenon, e.g., whether our solution is able to quantize and encode signals varying at $MHz$ level. The current design uses COTS components for multiplexers, comparators, etc.; it is of the authors' opinion that designing a custom Integrated Circuit~(IC) can enhance the performance in terms of power consumption, frequency tolerance, etc., which is out of the scope of this work. Thirdly, the power consumption of the proposed encoding circuit is $\approx 24~\mu \rm{W}$ (excluding the transmission power). As mentioned previously, it might be necessary to further bring this power down as energy-harvesting techniques are only able to produce tens of $\mu \rm{W}$~\cite{Khan2018} to power the entire sensing/transmitting device including transmission.

\section{Conclusion and Future Work}\label{sec:conc}
A novel modular sensing architecture that enables high-density persistent wireless monitoring has been proposed for Wireless Sensor Networks~(WSNs) that separates sensing and computational aspects between low-power, low-cost analog substrate and digital nodes of a traditional WSN. In order to achieve low-power (persistent sensing) and low-cost (high-density sensing) objectives, the analog sensors have been equipped with Analog Joint Source Channel Coding~(AJSCC) capabilities realized via Metal Oxide Semiconductor Field Effect Transistor~(MOSFET) characteristics. Techniques have been proposed for sensor encoding, decoding and AJSCC parameter optimization. The proposed techniques have been evaluated via MATLAB and LTSpice simulations which indicate that the proposed solution meets the needs of high-density, persistent wireless monitoring applications.

As part of the future work, (i)~we will first study the above discussed issues; specifically, we will investigate methods to further reduce the amount and effect of quantization; we will also study the frequency response of our circuit, e.g., finding the maximum frequency of the sensing signal supported by our circuit; (ii)~regarding the power consumption, we can notice in Fig.~\ref{fig:rmse_phi}, for $\phi<0.4~\rm{V}$, that the RMSE is non-zero for both $V_{gs}$ and $V_{ds}$. This suggests that having more than $10$ curves (i.e., AJSCC levels) in a single MOSFET will degrade the RMSE. To alleviate this undesired behavior, a multi-MOSFET architecture can be adopted. For example, in case $20$ AJSCC levels are desired, we can have four MOSFETs whose $V_{gs}$ values/curves are interwined so that there are only $5$ curves in each MOSFET, and $20$ combining all four. This achieves $\phi=0.2~\rm{V}$ without degradation in RMSE, unlike what we observe in Fig.~\ref{fig:rmse_phi}. All these four MOSFETs will only need one stage precircuit; and one precircuit can be reused for all four MOSFETs as only one of them is ON at a time. This reduces power consumption to $\approx 8~\mu\rm{W}$, making the circuit ultra low power. Additionally, it is possible that the performance of the MOSFET encoding varies with temperature. To compensate for this undesired behavior, the above multi-MOSFET architecture can again be leveraged---e.g., consider two MOSFETs with opposing temperature sensitivities---so that the temperature sensitivity will be canceled in their combination. In addition, we will investigate the following as part of our future work: (iii)~study our MOSFET-based encoding and decoding under realistic wireless channel conditions; (iv)~upon the success of (iii), conduct a pilot study of underwater pollution monitoring using our architecture, where we will investigate the possibility of designing our sensors including MOSFETs with biodegradable materials; (v)~we will investigate smart home/health applications, where both environmental and health monitoring signals are considered.

\balance
\bibliographystyle{IEEEtran}
\small
\bibliography{MyPublications_v1,career_v1,sensor_v5,references_v3.0,cross-layer_v3,underwater_v19,RobustConsensus,refs-vidya}

\begin{thebibliography}{10}
\providecommand{\url}[1]{#1}
\csname url@rmstyle\endcsname
\providecommand{\newblock}{\relax}
\providecommand{\bibinfo}[2]{#2}
\providecommand\BIBentrySTDinterwordspacing{\spaceskip=0pt\relax}
\providecommand\BIBentryALTinterwordstretchfactor{4}
\providecommand\BIBentryALTinterwordspacing{\spaceskip=\fontdimen2\font plus
\BIBentryALTinterwordstretchfactor\fontdimen3\font minus
  \fontdimen4\font\relax}
\providecommand\BIBforeignlanguage[2]{{%
\expandafter\ifx\csname l@#1\endcsname\relax
\typeout{** WARNING: IEEEtran.bst: No hyphenation pattern has been}%
\typeout{** loaded for the language `#1'. Using the pattern for}%
\typeout{** the default language instead.}%
\else
\language=\csname l@#1\endcsname
\fi
#2}}

\bibitem{Sadhu2017wons}
V.~Sadhu, X.~Zhao, and D.~Pompili, ``{Energy-efficient analog sensing for
  large-scale, high-density persistent wireless monitoring},'' in \emph{IEEE
  Annual Conference on Wireless On-Demand Network Systems and Services (WONS)},
  2017.

\bibitem{Sadhu2018ucomms}
V.~Sadhu, S.~Devaraj, and D.~Pompili, ``{Energy-efficient Wireless Analog
  Sensing for Persistent Underwater Environmental Monitoring},'' in \emph{IEEE
  Underwater Communications and Networking Conference (UComms)}, aug 2018, pp.
  1--4.

\bibitem{Rahmati2019Secon}
M.~Rahmati, V.~Sadhu, and D.~Pompili, ``{ECO-UW IoT: Eco-friendly Reliable and
  Persistent Data Transmission in Underwater Internet of Things},'' in
  \emph{Annual IEEE International Conference on Sensing, Communication, and
  Networking (SECON)}, Boston, MA, USA, jun 2019, pp. 1--9.

\bibitem{Lu2014}
N.~Lu, N.~Cheng, N.~Zhang, X.~Shen, and J.~W. Mark, ``{Connected vehicles:
  Solutions and challenges},'' \emph{IEEE Internet of Things Journal}, vol.~1,
  no.~4, pp. 289--299, aug 2014.

\bibitem{Sadhu2016icac}
V.~Sadhu, G.~Salles-Loustau, D.~Pompili, S.~Zonouz, and V.~Sritapan, ``{Argus:
  Smartphone-enabled human cooperation via multi-agent reinforcement learning
  for disaster situational awareness},'' in \emph{IEEE International Conference
  on Autonomic Computing (ICAC)}, 2016.

\bibitem{Chen2014}
S.~Chen, H.~Xu, D.~Liu, B.~Hu, and H.~Wang, ``{A vision of IoT: Applications,
  challenges, and opportunities with China Perspective},'' \emph{IEEE Internet
  of Things Journal}, vol.~1, no.~4, pp. 349--359, aug 2014.

\bibitem{Sadhu2017percom}
V.~Sadhu, G.~Salles-Loustau, D.~Pompili, S.~Zonouz, and V.~Sritapan, ``{Argus:
  Smartphone-enabled human cooperation for disaster situational awareness via
  MARL},'' in \emph{IEEE International Conference on Pervasive Computing and
  Communications Workshops, PerCom Workshops}, 2017.

\bibitem{Kato2017}
N.~Kato, I.~Bisio, and J.~Liu, ``{Guest Editorial 'Things' as Intelligent
  Sensors and Actuators in the Users' Context: Processing and Communications
  Issues},'' \emph{IEEE Internet of Things Journal}, vol.~4, no.~2, pp.
  297--298, apr 2017.

\bibitem{Mahmud2017}
M.~S. Mahmud, H.~Wang, A.~M. Esfar-E-Alam, and H.~Fang, ``{A Wireless Health
  Monitoring System Using Mobile Phone Accessories},'' \emph{IEEE Internet of
  Things Journal}, vol.~4, no.~6, pp. 2009--2018, dec 2017.

\bibitem{Sadhu2019tmc}
V.~Sadhu, S.~Zonouz, V.~Sritapan, and D.~Pompili, ``{CollabLoc:
  Privacy-preserving Multi-modal Collaborative Mobile Phone Localization},''
  \emph{IEEE Transactions on Mobile Computing}, pp. 1--13, 2019.

\bibitem{Shannon49}
C.~Shannon, ``Communication in the presence of noise,'' \emph{Proceedings of
  the IRE}, 1949.

\bibitem{Hekland05}
F.~Hekland, G.~Oien, and T.~Ramstad, ``Using 2:1 {S}hannon mapping for joint
  source-channel coding,'' in \emph{Data Compression Conference (DCC)}, March
  2005, pp. 223--232.

\bibitem{Zhao2018a}
X.~Zhao, V.~Sadhu, A.~Yang, and D.~Pompili, ``{Improved Circuit Design of
  Analog Joint Source Channel Coding for Low-Power and Low-Complexity Wireless
  Sensors},'' \emph{IEEE Sensors Journal}, vol.~18, no.~1, pp. 281--289, jan
  2018.

\bibitem{Chao2011}
P.~C. Chao, ``{Energy harvesting electronics for vibratory devices in
  self-powered sensors},'' \emph{IEEE Sensors Journal}, vol.~11, no.~12, pp.
  3106--3121, 2011.

\bibitem{piezodatasheet}
``{Piezo Protection Advantage, Mide Technology Corporation},''
  \url{http://info.mide.com/hubfs/ppa-piezo-product-datasheet.pdf}.

\bibitem{Shah.etal.AHN2003}
R.~C. Shah, S.~Roy, S.~Jain, and W.~Brunette, ``Data mules: {M}odeling and
  analysis of a three-tier architecture for sparse sensor networks,'' \emph{Ad
  Hoc Networks}, vol.~1, no.~2, pp. 215--233, Sept. 2003.

\bibitem{Wang.etal.EJAiSP2003}
H.~Wang, D.~Estrin, and L.~Girod, ``Preprocessing in a tiered sensor network
  for habitat monitoring,'' \emph{EURASIP J. Advances in Signal Processing},
  vol. 2003, no.~4, pp. 1--10, Dec. 2003.

\bibitem{Garcia11}
J.~Garcia-Naya, O.~Fresnedo, F.~Vazquez-Araujo, M.~Gonzalez-Lopez, L.~Castedo,
  and J.~Garcia-Frias, ``Experimental evaluation of analog joint source-channel
  coding in indoor environments,'' in \emph{IEEE International Conference on
  Communications (ICC)}, June 2011, pp. 1--5.

\bibitem{Romero14}
S.~Romero, M.~Hassanin, J.~Garcia-Frias, and G.~Arce, ``Analog joint source
  channel coding for wireless optical communications and image transmission,''
  \emph{Journal of Lightwave Technology}, vol.~32, no.~9, pp. 1654--1662, May
  2014.

\bibitem{Saleh12}
A.~Abou~Saleh, W.-Y. Chan, and F.~Alajaji, ``Compressed sensing with nonlinear
  analog mapping in a noisy environment,'' \emph{IEEE Signal Processing
  Letters}, vol.~19, no.~1, pp. 39--42, Jan 2012.

\bibitem{Zhao2016}
X.~Zhao, V.~Sadhu, and D.~Pompili, ``{Low-power all-analog circuit for
  rectangular-type analog joint source channel coding},'' in \emph{IEEE
  International Symposium on Circuits and Systems (ISCAS)}, vol. 2016-July,
  2016.

\bibitem{Khan2018}
O.~Khan, A.~Niknejad, and K.~Pister, ``{Ultra low-power transceiver SoC designs
  for IoT, NB-IoT applications},'' in \emph{2018 IEEE Custom Integrated
  Circuits Conference (CICC)}.\hskip 1em plus 0.5em minus 0.4em\relax IEEE, apr
  2018, pp. 1--77.

\bibitem{Zhao2017}
X.~Zhao, V.~Sadhu, T.~Le, D.~Pompili, and M.~Javanmard, ``{Towards low-power
  wearable wireless sensors for molecular biomarker and physiological signal
  monitoring},'' in \emph{IEEE International Symposium on Circuits and Systems
  (ISCAS)}, 2017.

\bibitem{Sadhu2019iscas}
V.~Sadhu, S.~Devaraj, and D.~Pompili, ``{Towards Ultra-Low-Power Realization of
  Analog Joint Source-Channel Coding using MOSFETs},'' in \emph{IEEE
  International Symposium on Circuits and Systems (ISCAS)}, Sapporo, Japan, may
  2019, pp. 1--5.

\bibitem{randraw}
A.~Bar-Guy, ``{MATLAB - Efficient Random Variates Generator},''
  https://www.mathworks.com/matlabcentral/fileexchange/7309-randraw.

\end{thebibliography}

\vspace{-0.5cm}
\begin{IEEEbiography}[{\includegraphics[width=1in,height=1.25in,clip,keepaspectratio]{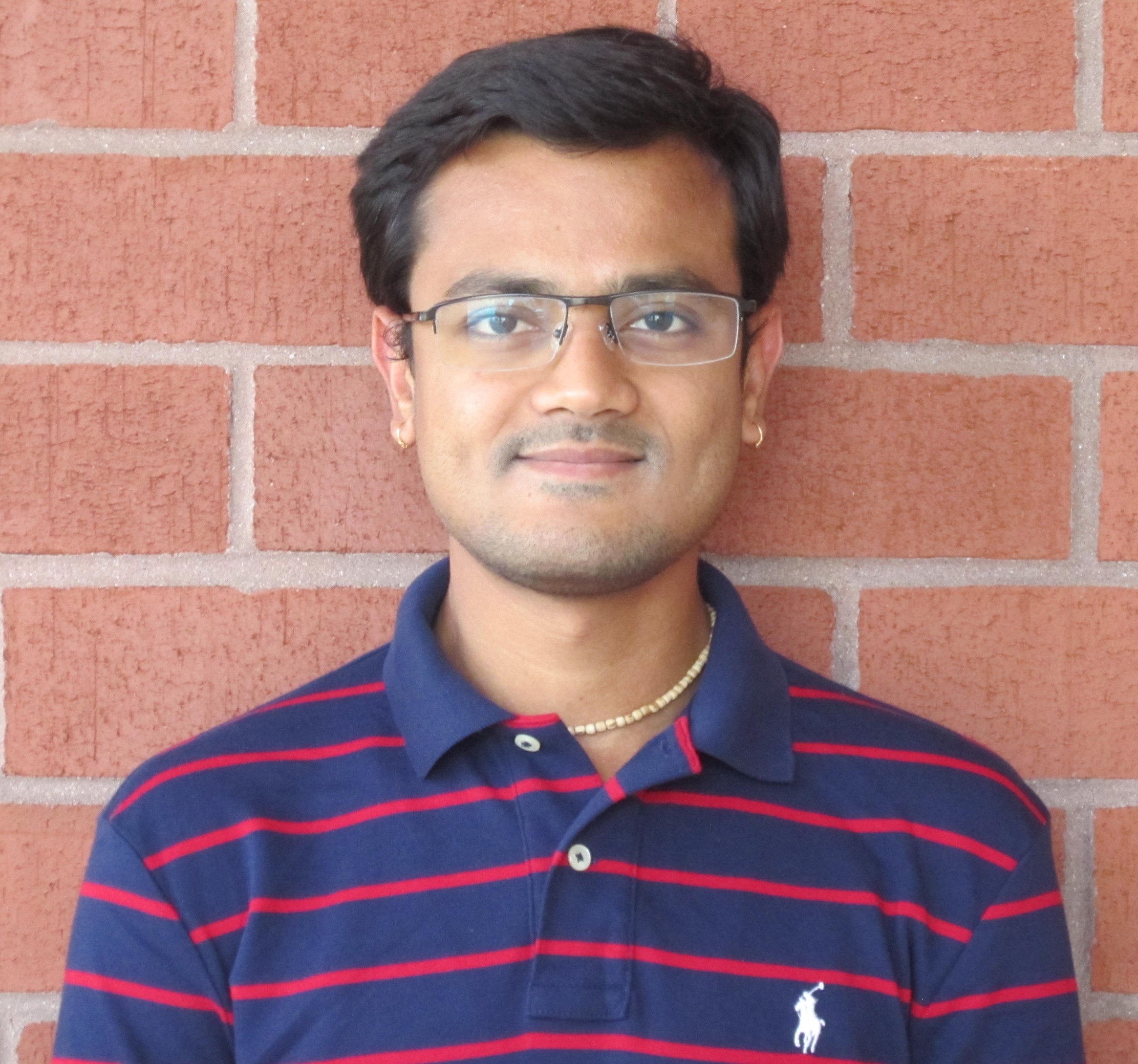}}]{Vidyasagar~Sadhu} (S'17) received the B.Tech. and M.Tech. degrees (under dual degree program) in Electrical and Computer Engineering~(ECE) from the Indian Institute of Technology, Chennai, India, in 2012. He worked in industry for two years before joining in 2014 the Ph.D. program in ECE at Rutgers University, where he works in the Cyber-Physical Systems Laboratory~(CPS~Lab) led by Dr.~Pompili. His research interests are in the domain of distributed computing and machine learning, mobile sensing, planning under uncertainty, and wireless sensor networks. He has contributed to several research projects, and is a co-recipient of the NSF Student Research Grant Award and of the Qualcomm Innovation Fellowship Finalist Award.
\end{IEEEbiography}

\vspace{-1.0cm}
\begin{IEEEbiography}[{\includegraphics[width=1in,height=1.25in,clip,keepaspectratio]{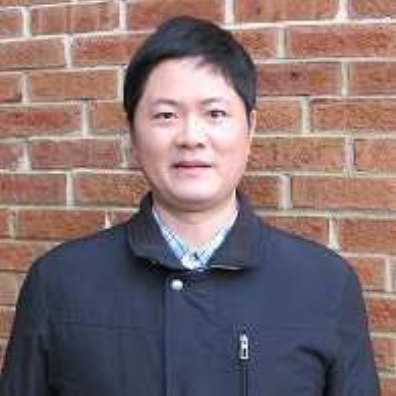}}]{Xueyuan~Zhao}
received the B.Eng. and M.Eng. degrees from the Beijing University of Posts and Telecommunications~(BUPT). He is pursuing the Ph.D. degree in ECE at Rutgers University, where he is a member of the CPS~Lab led by Dr.~Pompili. He worked in industry after graduation from BUPT, and contributed to multiple research and development projects. He is an inventor and co-inventor of a number of U.S. patents. He is a co-recipient of the Qualcomm Innovation Fellowship Finalist Award.
\end{IEEEbiography}

\vspace{-1.0cm}
\begin{IEEEbiography}[{\includegraphics[width=1in,height=1.25in,clip,keepaspectratio]{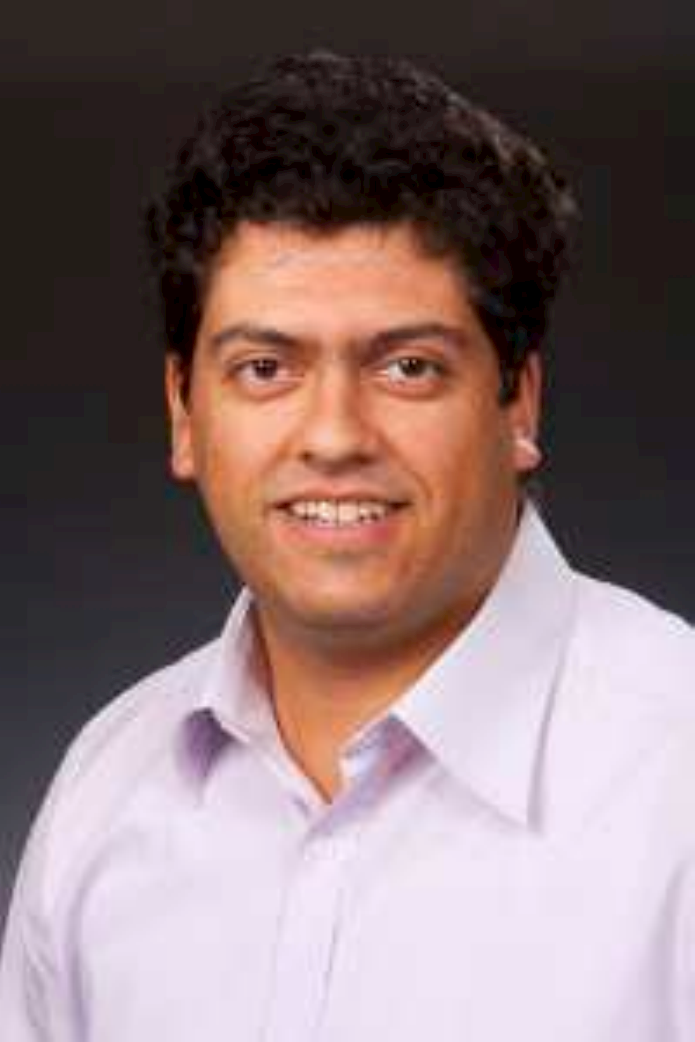}}]{Dario~Pompili} is an associate professor with the Dept. of ECE at Rutgers University. Since joining Rutgers in 2007, he has been the director of the CPS~Lab, which focuses on mobile edge computing, wireless communications and networking, acoustic communications, and sensor networks. He received his PhD in ECE from the Georgia Institute of Technology in 2007. He had previously received his `Laurea' (combined BS and MS) and Doctorate degrees in Telecommunications and System Engineering from the U. of Rome ``La Sapienza,'' Italy, in 2001 and 2004, respectively. He has received a number of awards in his career including the NSF CAREER'11, ONR Young Investigator Program'12, and DARPA Young Faculty'12 awards. In 2015, he was nominated Rutgers-New Brunswick Chancellor's Scholar. He served on many international conference committees taking on various leading roles. He published about 150 refereed scholar publications, some of which selected to receive best paper awards: with more than 10K citations, Dr.~Pompili has an h-index of 40 and an i10-index of 97 (Google Scholar, Mar'20). He is a Senior Member of the IEEE Communications Society~(2014) and a Distinguished Member of the ACM~(2019). He is currently serving as Area Editor for IEEE Transactions on Mobile Computing~(TMC).
\end{IEEEbiography}

\end{document}